\documentclass{jfm} 
\nonstopmode
\useAMSsubequations
\usepackage{newtxtext}
\usepackage{newtxmath}
\usepackage{amsfonts} 
\usepackage{amsmath}
\usepackage{amssymb}
\usepackage{bm}
\usepackage{natbib}
\usepackage[section]{placeins}
\usepackage{xcolor}
\usepackage{graphicx}
\usepackage{booktabs}
\usepackage{hyperref}
\hypersetup{
    colorlinks = true,
    urlcolor   = blue,
    citecolor  = black,
}
\usepackage{subcaption}
\usepackage{siunitx} 
\DeclareSIUnit{\pixel}{px}
\DeclareSIUnit{\px}{px}
\DeclareSIUnit{\frame}{frame}
\DeclareSIUnit\litre{l} 

\usepackage{physics}

\graphicspath{{},{./}}
\newlength{\figwidth}
\renewcommand{\emph}[1]{\textsl{#1}} 

\usepackage{todonotes} 
\usepackage{soul}
\sethlcolor{orange}

\newcommand{\todohl}[2]{%
    \makeatletter\if@todonotes@disabled%
    #1%
    \else%
    \hl{#1}\todo{#2}%
    \fi\makeatother%
}

\newcommand{\mucl}{\ensuremath{\mu_\text{cl}}}
\newcommand{\vcap}{\ensuremath{v_\text{c}}} 
\newcommand{\us}{{\ensuremath{u_s}}}
\newcommand{\un}{{\ensuremath{u_n}}}
\newcommand{\ux}{{u}} 
\newcommand{\uz}{{v}} 

\newcommand{\Lz}{{\ensuremath{\mathcal{L}_z}}}
\newcommand{\Lw}{{\ensuremath{\mathcal{L}_w}}}
\newcommand{\NLzz}{{\ensuremath{\mathcal{N}_{z}}}}
\newcommand{\NLww}{{\ensuremath{\mathcal{N}_{w}}}}

\newcommand{\Tdim}{{\ensuremath{ \tilde T }}}
\newcommand{\Zampdim}{{\ensuremath{ \tilde Z }}}
\newcommand{\Wampdim}{{\ensuremath{ \tilde W }}}
\newcommand{\Fampdim}{{\ensuremath{ \tilde F }}}
\newcommand{\Uampdim}{{\ensuremath{ \tilde U }}}
\newcommand{\lengthscale}{{\ensuremath{ \ell }}} 
\newcommand{\Zamp}{{\ensuremath{ Z }}}
\newcommand{\Wamp}{{\ensuremath{ W }}}
\newcommand{\Famp}{{\ensuremath{ F }}}

\newcommand{\eps}{{\ensuremath{\epsilon}}}
\newcommand{\epss}{{\ensuremath{\epsilon'}}}

\newcommand{\conjugate}[1]{{\ensuremath{ {#1}^* }}}

\newcommand{\evmat}{{\ensuremath{\mathbf{L}}}}
\newcommand{\evmatlinear}{{\ensuremath{\evmat_0}}}
\newcommand{\evmatsmall}{{\ensuremath{ \delta \evmat }}}

\newcommand{\cc}{{\text{c.c.}}}
\newcommand{\adv}{{\ensuremath{\partial_a}}} 
\newcommand*{\deq}{%
    \mathrel{\vbox{\offinterlineskip\ialign{%
        \hfil##\hfil\cr
        \scalebox{1.13}[1.06]{\textasciicircum}\cr
        \noalign{\kern-1.1ex}
        $=$\cr
    }}}}
\makeatletter
\newcommand*{\defeq}{\mathrel{\rlap{%
    \raisebox{0.3ex}{$\m@th\cdot$}}%
\raisebox{-0.3ex}{$\m@th\cdot$}}%
    =}
\newcommand*{\eqdef}{=\mathrel{\rlap{%
    \raisebox{0.3ex}{$\m@th\cdot$}}%
\raisebox{-0.3ex}{$\m@th\cdot$}}%
}
\makeatother

\definecolor{col_acoustic}{HTML}{0EB200}
\definecolor{col_mu}{HTML}{999900}
\definecolor{col_mucl}{HTML}{990000}
\definecolor{col_gammat}{HTML}{006699}
\definecolor{col_gamman}{HTML}{330099}
\definecolor{col_geo}{HTML}{777777}
\definecolor{col_nonres}{HTML}{444444}

\newcommand{\Cgat}[1]{\textcolor{black}{#1}}

\newcommand{\Cgeo}[1]{\textcolor{black}{#1}}

\colorlet{revision1color}{black}
\colorlet{revision2color}{black}
\newcommand{\revise}{\textcolor{revision1color}}
\newcommand{\rerevise}{\textcolor{revision2color}}
\newcommand{\revisecross}[1]{}

\title{Dancing rivulets in an air-filled Hele-Shaw cell}

\author{Grégoire Le Lay\aff{1}
\corresp{\email{gregoire.lelay@proton.me}},
    \and Adrian Daerr\aff{1}
}
\affiliation{\aff{1}Laboratoire Matière et Systèmes Complexes (MSC, UMR 7057), Université Paris Cité | CNRS, 75013 Paris, France}

\nonstopmode

\begin{document}
\maketitle

\begin{abstract}
    We study the behaviour of a thin fluid filament (a rivulet) flowing in an air-filled Hele-Shaw cell.
    Transverse and longitudinal deformations can propagate on this rivulet,
    although both are linearly attenuated in the parameter range we use.
    On this seemingly simple system,
    we impose an external acoustic forcing, homogeneous in space and harmonic in time.
    When the forcing amplitude exceeds a given threshold,
    the rivulet responds nonlinearly, adopting a peculiar pattern.
    We investigate the ``dance'' of the rivulet both experimentally using spatiotemporal measurements,
    and theoretically using a model based on depth-averaged Navier-Stokes equations.
    The instability is due to a three-wave resonant interaction between waves along the rivulet,
    the resonance condition fixing the pattern wavelength.
    Although the forcing is additive, the amplification of transverse and longitudinal waves is effectively parametric,
    being mediated by the linear response of the system to the homogeneous forcing.
    Our model successfully explains the mode selection and phase-locking between the waves,
    it notably allows us to predict the frequency dependence of the instability threshold.
    \revise{The dominant spatiotemporal features of the generated pattern
    are understood through a multiple-scale analysis.}
    \revisecross{Several questions remain open, we conclude by suggesting potentially fruitful future research directions.}
\end{abstract}

\begin{keywords}
    Nonlinear instability, Capillary waves, Parametric instability, Liquid bridges
\end{keywords}

\section{Introduction}
\label{sec:intro}

The study of instabilities has been a driving force in fluid mechanics
for the last two hundred years, \revise{and is part of virtually every
  modern course in fluid mechanics. Understanding the nature of
  various observed instabilities is the basis for their control, be it
  with the aim of suppressing or on the contrary exploiting them. This
  has led to the distinction of several archetypes of instability
  scenarios, which the instability described in this paper combines in
  a way that, as we discuss below, is uncommon yet likely relevant in
  other systems. Specifically the present system generalizes
  Faraday-type parametric instabilities to the systems where two types
  of waves coexist.}

In this paper, we present an experimental report and theoretical model
for a recently discovered instability that arises when a liquid
rivulet in air-filled Hele-Shaw cell is forced
acoustically~\citep{lelay2025}. A thin \revise{gravity-driven}
filament of liquid in a Hele-Shaw cell usually flows down vertically,
but when a spatially invariant horizontal forcing is applied, it
destabilizes and forms a complex spatio-temporal pattern. This pattern
has a finite wavelength, while the forcing is spatially homogeneous,
and it can be seen as a phase-locked superposition of transverse and
longitudinal waves propagating on the rivulet. \revise{Explaining
  the mechanism behind this pattern-forming instability and its consequences is the main motivation of this article}

In dynamical systems theory, it is common to distinguish
\revisecross{\textsl{external}, or }\textsl{additive,} forcing, and
\textsl{parametric}, or multiplicative, forcing. When forcing a system
additively, one usually expects a linear response composed of the same
space spatial and temporal frequency components as the excitation. On
the contrary, exciting a system parametrically leads to more complex,
nonlinear behaviour: this is notably illustrated in fluid dynamics by
the much-studied Faraday instability~\citep{faraday1831, douady1990,
  kumar1994, bongarzone2022} where the control parameter is the
gravitational acceleration.
Indeed, vibrating a liquid vertically will famously leads to secondary
instabilities~\citep{fauve1991, daudet1995, tufillaro1989}, complex
interactions between modes~\citep{residori2007}, partially stationary
patterns when the forcing is localized~\citep{moisy2012}, intricate
spatiotemporal patterns~\citep{edwards1994}, mean
flows~\citep{guan2023}, it can also prevent coalescence of
droplets~\citep{couder2005}, or can even be used to parametrically
stabilize unstable fluid configurations~\citep{apffel2020}.

In our system we apply an \revisecross{external, }additive, acoustic forcing.
This forcing, which is homogeneous in space and harmonic in time, leads to a linear response of the rivulet with the same spatiotemporal characteristics.
However, this linear response effectively behaves as a multiplicative parameter coupling longitudinal and transverse waves on the rivulet.
When the forcing is strong enough, this leads to the seemingly paradoxical parametric destabilization of a system that is forced additively.
This original mechanism allows an initially homogeneous rivulet to display a pattern presenting a well-defined wavelength although the forcing does not depend on space.

Moreover, this instability relies on the co-amplification of two waves that cooperate by interacting constructively instead of competing for the available energy.
This constructive interaction takes the form of a triadic resonant nonlinear interaction between the two co-amplified waves and the linear response to forcing.
Such an interaction is reminiscent of the case of Faraday waves in a
Hele-Shaw cell~\citep{douady1989, li2019, bongarzone2023}, \revise{and
  of other physical systems where a parametric coupling between waves
  allows for their amplification: for example }internal waves in a
variable-depth container~\citep{Szoeke1983, Benilov1987}, the
elliptical instability in rotating flows~\citep{kerswell2002,
  bars2015, lemasquerier2017, lebars2020}, or even, in a different
field outside of fluid mechanics, the biphoton mode in optical
parametric oscillators~\citep{amon2009} ; with the notable difference
that contrary to the previously cited example, in our system the
co-amplifying waves are different in nature, having distinct
propagation mechanisms and thus very different dispersion relations.
\revise{Discussing the emergence of a parametric instability as a
  limit case of triadic wave interaction where the forcing acts by
  producing a response in the form of a wave with
  zero wavenumber, provides a second
  motivation for this work.}

In the system we study, we inject a small stream of liquid in between
two close by, parallel and vertical glass plates (forming a Hele--Shaw
cell). Since the liquid we use perfectly wets the glass, it forms a
liquid bridge spanning the cell thickness, delimited by two
semicircular menisci. \revisecross{The plates being vertical, this
  bridge falls down }Under the influence of gravity,
\revisecross{forming}\revise{the liquid forms }a thin elongated
filament \revise{along the entire height of the cell}, henceforth
termed a \textsl{rivulet}.

\revisecross{The rivulet behaviour is subject to the nonlinear interactions between the fluid flow inside the bridge
and the geometry of its free surface, which can often lead to complex behaviour
Because of gravity, a downward base flow always takes place inside the rivulet.
Due to the reduced spacing between the walls of the cell,
the viscous fluid adopts}\revise{The gravity-driven vertical base
flow displays a Poiseuille velocity profile with zero slip on the
bounding glass plates.}
At low flow rate this base state is stable, and it destabilizes when the flow rate exceeds a critical value $Q^*$~\citep{daerr2011}.
In this study, we will only consider the case $Q < Q^*$ where the straight rivulet of constant width is linearly stable.

Capillarity acts as a restoring force towards the straight rivulet state, allowing waves to propagate.
These small deformations propagate in the reference frame advected downwards at the mean flow speed inside the rivulet~$u_0$.
Waves on the rivulet are systematically damped by viscosity, which acts on two different level:
in the bulk of the rivulet, and near the edges of the menisci when they are displaced.
Indeed, moving the rivulet transversally at a finite speed implies displacing the fluid layer of vanishing thickness near the menisci edges,
which induces important dissipation.

In order to excite the system, we apply a pressure gradient between the right and left sides of the cell,
generating a transverse force on the rivulet that does not depend on the vertical coordinate.
The rivulet responds linearly to this forcing,
and displays a pattern-forming instability when the excitation amplitude is above a certain threshold.

The main purpose of this paper is to describe experimentally and explain theoretically the features of this instability,
the conditions under which it can develop, and the way it saturates.
We start in section~\ref{sec:waves} by explaining our experimental method, we recall the physics ingredients at play,
and we describe the waves that can propagate on a rivulet.
In section~\ref{sec:pearl-necklace-instability}, we provide a qualitative experimental description of the phenomenon,
we show how parametric coupling waves leads to amplification of perturbations,
and we identify the mathematical resonance condition that selects the pattern.
In section~\ref{sec:pattern} we then analyse the pattern growth in detail,
calculating the excitation threshold, the relative amplitudes of the interacting waves, the saturation mechanism,
and the maximum forcing before rivulet break-up.
In the conclusive section~\ref{sec:conclusion} we revisit experimental observations in the light of our analysis,
before providing a brief summary and presenting several promising future research directions.

\section{Fluid rivulets}
\label{sec:waves}

\subsection{Experimental set-up}
\label{subsec:setup}

\begin{figure}
    {\includegraphics[scale=1]{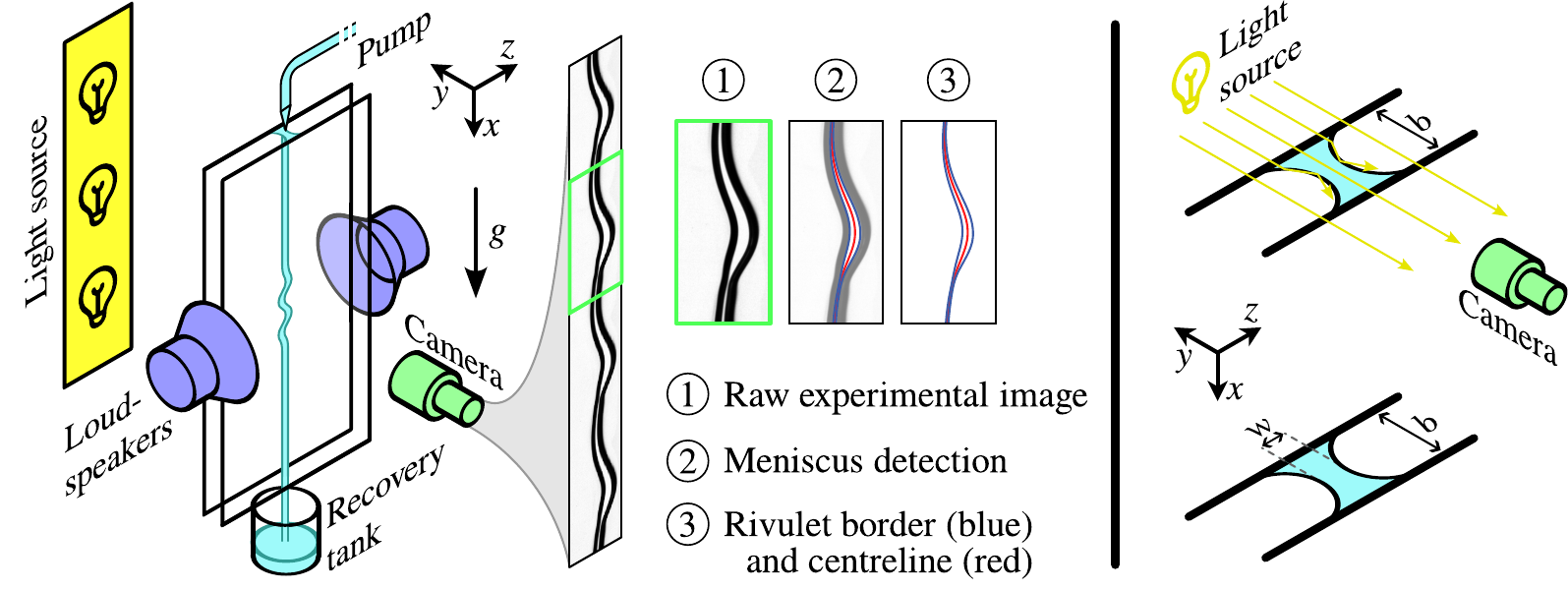}}%
    \caption{(left) Experimental apparatus (not to scale, schematic view).
    The measuring scale and the pump are not shown.
    \\
    (middle) Typical image, with detection of path and width of the rivulet.
    The path is defined as the centreline of the bright zone, which corresponds to the interior of the rivulet.
    The width is the distance between the detected menisci (blue lines in the figure)
    \\
    (right) On top, illustration of the fact that light refracted by the menisci does not reach the camera.
    On the bottom, definition of the cell spacing $b$ and rivulet width $w$.
    }
    \label{fig:schema}
\end{figure}

A schematic diagram of the experimental set-up is shown in figure~\ref{fig:schema}.
The Hele-Shaw cell consists of two parallel float glass plates of dimensions
$\SI{1}{\meter}\times\SI{10}{\centi\meter}\times\SI{6}{\milli\meter}$ forming an
air-filled gap $b$  \SIrange[range-units=single]{.5}{.6}{\milli\meter} thick depending on the experiments.
The gap thickness is imposed by mylar (PET) spacers of known thickness.
PTFE oil is injected locally at the top of the vertically set cell,
through a syringe tip fed by a gear pump from the main liquid reservoir.
Under the cell, a recovery tank placed on a lab scale collects the oil.
By reading the weight of the tank as a function of time,
we are able to recover the flow rate $Q$ that we inject in the cell.
The recovery tank is automatically flushed to the main reservoir whenever its weight exceeds a given threshold.
On opposite sides of the cell, two loudspeakers (Monacor SP-60/4) are
connected to the air gap via adapter plates that seal the speaker
front save
for a small opening.
The speakers are in anti-parallel configuration (\emph{push-pull}), the
movement of the membranes being antisymmetric with respect to the cell centre-line.

The liquid we used (PTFE oil Galden HT135, density
$\rho=\SI{1.71}{\gram\per\milli\litre}$, surface tension
$\gamma = \SI{17}{\milli\newton\per\meter}$, kinematic viscosity
$\nu = \SI{1}{\milli\meter\squared\per\second}$) meets the glass at a
vanishing contact angle, measured in a drop spreading experiment to be
smaller than \SI{0.5}{\degree}.
The oil therefore readily wets both
plates of the Hele-Shaw cell, and forms a liquid bridge between them.
A \si{\centi\metre}-size obstacle near the injection point breaks the
initial jet, reducing the fluid velocity to near zero independent of
the injection diameter, and ensuring wetting.
Under the action of gravity, the oil flows around the obstacle and
detaches from its lower tip to form a continuous
liquid stream spanning the entire cell height (\SI{1}{\meter}),
henceforth termed \emph{rivulet}.
This thin liquid filament is
delimited by two menisci on the sides, between which the falling fluid
flows downwards.
Near the top of the cell the rivulet thins as the fluid accelerates,
and reaches an equilibrium width when friction forces balance gravity.
\revise{%
The equilibrium flow speed inside the rivulet is noted $u_0 = g /\mu$, where $g$ is the acceleration of gravity,
and $\mu = 12\,\nu/b^2$ is a friction coefficient which depends on the velocity profile in the rivulet
(here a parabolice Poiseuille flow is assumed).}
The equilibrium flow speed is attained at a distance $u_0/\mu = g b^4/(12\nu)^2 \approx \SI{10}{\milli\meter}$
from the injection site.
All the measurements presented in this study are made below this point,
where the rivulet width is assumed to have reached its equilibrium value $w_0$.

The positions of the menisci bordering the rivulet are readily
determined experimentally.
In order to visualize these borders with high contrast,
we position the camera normal to the plates, and a bright back
lighting panel at a distance much larger than the field of view.
The light of the LED panel passes through the air and the (transparent) oil,
but it is refracted at the menisci.
The rivulet borders thus appear black on a bright background on experimental images.

Most of the experimental images are obtained using a gigabit-Ethernet industrial camera (AVT Manta G-223 B\revise{, max framerate \SI{400}{\hertz}, full resolution \SI[parse-numbers=false]{1024\times2088}{px} i.e. \SI{2}{Mpx}})
controlled by the open-source software \emph{Limrendir}.
When recording periodic phenomena,
we use a well-controlled frequency mismatch between the camera frame rate and the excitation period.
This technique allows us to visualize the rivulet with high temporal resolution,
even when the excitation frequency exceeds the maximum frame rate of the camera.
When true high time resolution is needed (i.e.\ when visualizing non-periodic phenomena),
we employ a high-speed camera (Chronos 1.4 \SI{8}{GB}\revise{, max. framerate \SI{20}{\kilo\hertz}, max. resolution \SI[parse-numbers=false]{1080\times1920}{px} i.e. \SI{2}{Mpx}}).
\revise{For the experiments presented in this paper, we used forcing frequencies between 10 and 2000~Hz.
The sampling frequency is always chosen so that we have more than 10 images per forcing period.}

The videos acquisitions are exploited using in-house Python routines,
which automatically detect the position $z(x,t)$ and width $w(x,t)$ of the rivulet.
The position of the rivulet corresponds to the centerline of the liquid bridge,
while its width is defined as the distance between the two menisci.
\revise{We proceed as follows:
    (i) the image is filtered (gaussian smoothing, vignetting correction, contrast optimization by normalisation),
    (ii) for each (horizontal) line of the image, the two local minima of luminosity are associated with the menisci borders,
    and the zone between these minima is the rivulet interior; the
    minimum value in this zone is subtracted from the
    luminosity in the following computations
    (iii) we estimate the position of the rivulet by computing the barycentre of the luminosity inside the rivulet,
    and the width by computing the full width at half maximum (FWHM) of the luminosity profile in the rivulet interior.}
These two quantities, position and width, correspond to different kind of perturbations of the rivulet base state.
The reference situation corresponds to a straight rivulet of constant width flowing vertically.
A non-zero position corresponds to a transverse perturbation of the rivulet,
while a width different from the rest width $w_0$ corresponds to a longitudinal perturbation.

\subsection{Dynamical equations}
\label{subsec:equations}

In this section we establish the mathematical model we use to describe the evolution of the rivulet.
Throughout this study we assume that the rivulets remain \emph{slender} \revise{in the $(x,z)$ plane},
meaning that their position $z$ and width $w$ vary on scales that are large compared to the base width $w_0$ of the rivulet.
Mathematically, this translates as $\abs{\partial_x z} \ll 1$ and $\abs{\partial_x w} \ll 1$.

In order to describe the dynamics of the system we start from first principles,
i.e.\ the Navier-Stokes equations.
We perform a depth-averaging (along the $y$ direction, normal to the plates),
assuming a parabolic profile (corresponding to a Poiseuille flow) for the velocity.
This laminar flow hypothesis is largely relevant,
as the typical Reynolds number associated with the flow inside the rivulet is of order $b u_0 / \nu \approx 150$.
We then integrate the result along the direction transverse to the rivulet path.
After dividing by the liquid density $\rho$,
we obtain the following equations for the depth-averaged fluid velocity $\vb{u}=\ux\,\vb{e}_x + \uz\,\vb{e}_z$:
\begin{align}
    \label{eq:nsvectoriel}
    w\,(\partial_t + \beta\vb{u}\cdot\grad) \vb{u}
    =& w\,\vb{g} - w\,\mu\,\vb{u}
    +  w\,\Gamma\,\grad\kappa_w
    + \qty(\Gamma\,\kappa_z - \mucl\,\vb{u}\cdot\vb{n} + \Pi)\,\vb{n}
\end{align}
where $\vb{n}$ is a unit vector normal to the rivulet path $z(x,t)$.
We also take into account mass conservation
\begin{align}
    \label{eq:massconsvectoriel}
    (\partial_t + \vb{u}\cdot&\grad) w = -w\,\grad \cdot \vb{u}
\end{align}
where $w(x,t)$ is the width of the rivulet.
The system is closed using the kinematic boundary condition $\uz = (\partial_t + \vb{u}\cdot\grad)z$.

The dynamical equation~\eqref{eq:nsvectoriel} represent the competition between several physical effects:
inertia, gravity, internal friction, streamwise pressure gradients inside the rivulet,
and transverse pressure forces\revisecross{ from outside the rivulet}.

The leftmost, inertial term features the numerical prefactor $\beta \simeq 1$ accounting for the velocity's $y$-profile.
As~\citet{gondretRabaud1997}, we set this prefactor to one in the following,
as it has only minor effects.
The internal friction term originates from the viscosity of the fluid and is given by Darcy's law,
with $\mu = 12\,\nu/b^2$ (assuming a parabolic velocity profile).
The streamwise pressure gradient term is due to inhomogeneities of
Laplace pressure along the path of the rivulet when its width is uneven.
It acts as a restoring force towards the constant-width situation.
This mechanism is at the origin of the propagation of longitudinal capillary waves along the rivulet,
as discussed in the next section.
Here $\Gamma = \pi\,\gamma/(2\,\rho)$, incorporating the $\pi/4$ corrective factor from~\citet{parkHomsy84} ;
and $\kappa_w$ is the curvature of the width profile of the rivulet $w(x,t)$
(at first approximation, $\kappa_w \approx \partial_{xx}w$).

The \revisecross{external }\revise{transverse }pressure forces normal to the interface are directed along $\vb{n}$.
There are three distinct mechanisms which generate a pressure gradient transverse to the rivulet:
\begin{itemize}
    \item The first contribution is the Laplace pressure due to the effective curvature of the path $\kappa_z  \approx \partial_{xx}z $.
    Indeed, when the rivulet is curved,
    a restoring force acts to straighten it back.
    This force originates in the \revise{change of }Laplace pressure \revise{due to the deformation of the menisci }caused by the curvature of the rivulet path~\citep{parkHomsy84}.
    This mechanism is at the origin of the propagation of transverse capillary waves along the rivulet,
    as discussed in the next section.
    \item The second contribution corresponds to the stress due to the important dissipation at the moving contact line.
    This term translates the fact that moving the rivulet in the direction normal to its path is energetically costly,
    since it involves the displacement of a vanishingly thin film on the meniscus border.
    The additional dissipation associated with transverse movement
    depends notably on the height of the film already present on the
    glass plates ahead of the moving meniscus.
    This film being the result of previous wetting events,
    rigorously, its local height depends the history of rivulet
    movement.
    We make the simplifying assumption that the rivulet is
    sliding on a film of constant effective thickness, neglecting
    potential spatial or temporal variations of the
    contact line friction coefficient $\mucl$.
    The effective depth $h$ is estimated to be about several
    micrometers, in accordance with recent measurements on this system~\citep{lelay2025film}.
    \item Last, the $\Pi$ term corresponds to the transverse force induced by a pressure difference $\pm\frac12 \rho \Pi$ between the two halves of the cell separated by the rivulet.
    The rivulet acts as a membrane separating the left and right sides of the cell in an airtight manner:
    this means it is sensitive to pressure differences between the air masses on its right and left sides.
    Such a pressure difference drives the rivulet transversally.
    Using loudspeakers at the side of the cell, we are able to impose an arbitrary pressure difference $\Pi(t) \neq 0$.
\end{itemize}

\subsection{Wave propagation on rivulets}
\label{subsec:waves}

In the absence of forcing, the rivulet is in its base state,
corresponding to a straight downwards flowing filament of constant width:
$z = z_0 = 0$, $w = w_0$ and $\vb{u} = u_0\,\vb{e}_x$ with $u_0  =g/\mu$.
Let us now consider a perturbation to this state, and see how it evolves.

\revise{We introduce a small nondimensional number $\eps \ll 1$ in the problem},
so that we can develop all the dynamic variables in successive powers of this small parameter,
in the fashion $a = a_0 + \eps a_1 + \eps^{2}a_2$ \ldots{} for $a=z, w, \ux, \uz, $ \ldots{}
\revise{The perturbation amplitude of the mid-surface position, for example, is then $\eps z_1$,
and the fact that $\eps \ll 1$ implies that this perturbation amplitude is small compared to the typical wavelength $\lambda$ of the perturbation.}

Moreover in the following we make the assumption of low damping,
that is of small $\mu$ and $\mucl$ of order $\eps^2$.
This assumption allows us to focus on the undamped propagation and
interaction of waves.
\revise{Since $\eps \ll 1$, this means that we are considering phenomena with typical frequencies $\omega$ that are very large compared to the damping frequencies $\mu$ and $\mucl/w_0$.
Note that despite the small damping assumption, we still consider the base flow speed $u_0$ to be of order unity.}

The linear (first order in $\eps$) approximation of the main equations
\eqref{eq:nsvectoriel} and \eqref{eq:massconsvectoriel} reads
\begin{subequations}\label{eq:massconsordreun}
\begin{align}
    w_0(\partial_t + u_0\partial_x) \uz_1
    =&
    \Gamma\, \partial_{xx}z_1
    \\
    w_0(\partial_t + u_0\partial_x) \ux_1
    =&
    w_0\,\Gamma\,\partial_{xxx}w_1
    \\
    (\partial_t + u_0\partial_x) w_1 =& -w_0 \partial_x \ux_1
\end{align}
\end{subequations}
Using the kinematic condition
$\uz_1 = (\partial_t + u_0\partial_x)z_1$ to eliminate $\uz_1$, this system can be
recast in terms of two linear evolution operators $\Lz$ and $\Lw$
\begin{subequations}
\begin{align}
    0 &= \qty ((\partial_t + u_0\partial_x)^2
    -
    \vcap^2 \partial_{xx}) z_1 \eqdef \Lz z_1
    \label{eq:dispersionequation_z}
    \\
    0 &= \qty( (\partial_t + u_0\partial_x)^2 + \vcap^2 {w_0}^2 \partial_{xxxx} )w_1 \eqdef \Lw w_1
    \label{eq:dispersionequation_w}
\end{align}
\end{subequations}
where we introduced the \emph{capillary velocity} $\vcap = \sqrt{\Gamma/w_0}$.
This velocity can be interpreted as the displacement speed of purely capillary transverse deformations along the rivulet at rest.
It corresponds to the phase speed of the sinuous waves in the reference frame of the falling liquid.

The propagation operators $\Lz$ and $\Lw$ correspond to two distinct types of waves that can exist on the surface of the rivulet.
These waves are linearly independent of one another.
The $z$-waves correspond to transverse waves, i.e.\ deformations of the path followed by the rivulet.
The $w$-waves correspond to longitudinal waves, i.e.\ modulations of the rivulet width.
These waves are also referred to as sinuous and varicose modes in the
literature.

\begin{figure}
    {\includegraphics[scale=1]{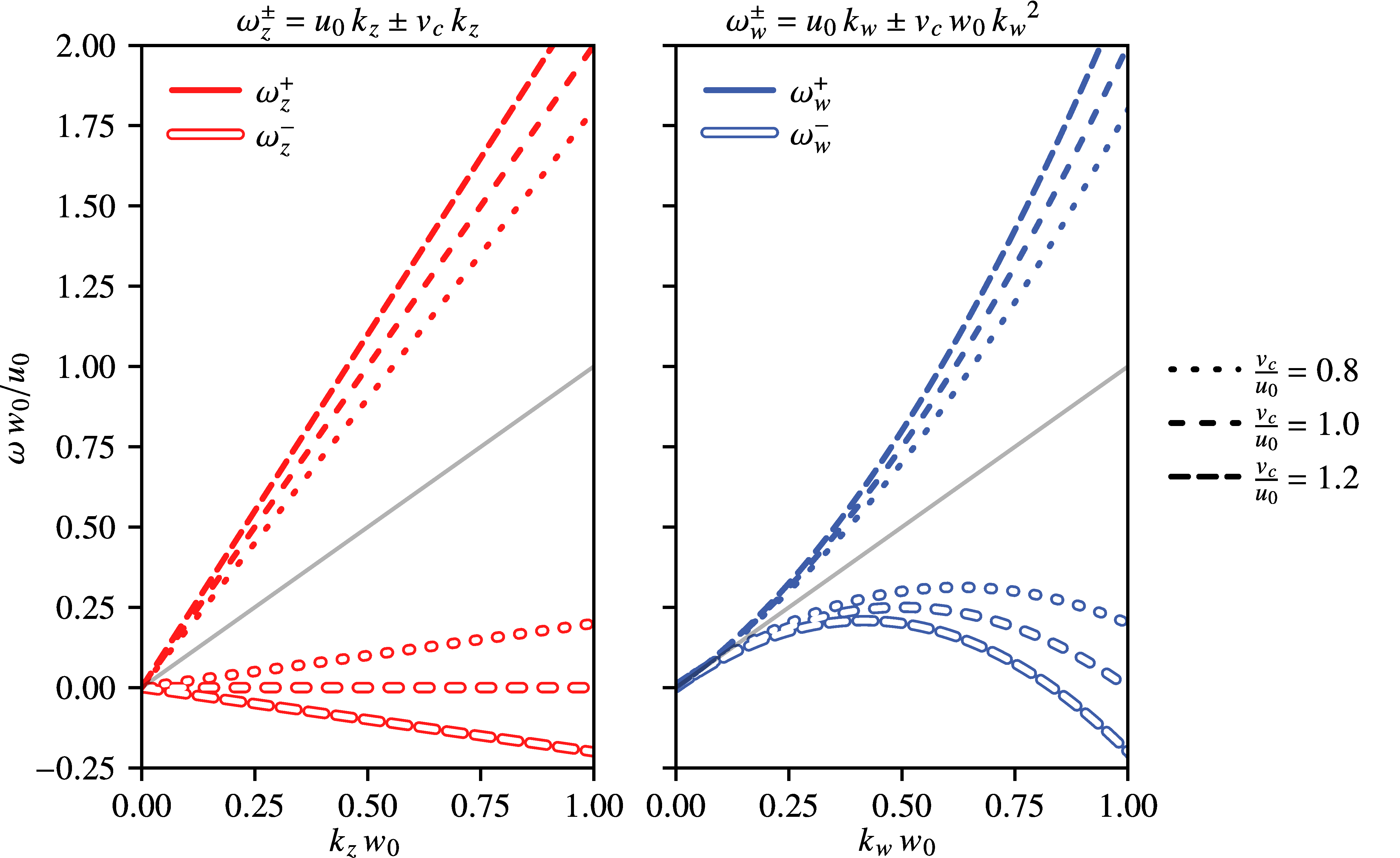}}
    \caption{Dispersion relations of transverse (left) and longitudinal waves (right)\revise{, in the absence of damping}.
    The axis is made dimensionless through appropriate scaling.
    The horizontal axes correspond to the dimensionless wavevector amplitude $k\, w_0$,
        while vertical axes correspond to the dimensionless angular frequency $\omega \, w_0/u_0$.
        The gray lines correspond to pure advection $\omega = k\, u_0$.
    }
    \label{fig:dispersionrelation_plot}
\end{figure}

Each type of wave supports two modes, corresponding to the dispersion relations
\begin{subequations} \label{eq:dispersionrelation}
\begin{align}
    \omega_{z}^{\pm} &= (u_0 \pm \vcap) k_z
    \label{eq:dispersionrelation_z}
    \\
    \omega_{w}^{\pm} &= u_0 k_w \pm \vcap w_0 {k_w}^2
    \label{eq:dispersionrelation_w}
\end{align}
\end{subequations}
which are plotted on figure~\ref{fig:dispersionrelation_plot}.

The transverse waves propagate non-dispersively,
while the dispersion relation of the longitudinal waves display a quadratic term.
For both waves, two branches coexist: one \emph{fast} branch for which the phase speed in the laboratory reference frame is faster than the flowing speed $u_0$,
and one \emph{slow} branch which displays phase speeds in the laboratory reference frame inferior to $u_0$ (and which can be negative).
These branch correspond respectively to positive (for fast waves) or negative (for slow waves) phase speed
in the reference frame advected with the flow speed $u_0$.

\section{Instability mechanism}
\label{sec:pearl-necklace-instability}

In the absence of forcing, the rivulet flows straight down, vertically,
and since the flow rate is always under the spontaneous meandering threshold $Q^*$, one observes that this base state is very robust.
This is because both transverse and longitudinal waves are linearly damped
(as can be seen when looking at the dynamical equations presented on section~\ref{subsec:resonance_condition} and onwards).
Experimentally, if a perturbation of any kind is imposed on the rivulet,
its amplitude is observed to decrease exponentially within a short timescale ($< \SI{100}{\milli\second}$) \revise{and over a short spatial distance ($< \SI{3}{\centi\meter}$)}.

However, while both transverse and longitudinal waves are linearly attenuated,
we discovered recently~\citep{lelay2025} that when a homogeneous, harmonic, acoustic forcing is applied,
the base state becomes unstable and the rivulet exhibits a distinct pattern combining both types of waves.
\revise{The birth of this pattern can be observed on Supplementary Movie SM2.}
The main features of this pattern is that while transverse and longitudinal perturbations travel at different speed,
they both share a commun spatial wavelength.
This should come as a surprise, as since the excitation is homogeneous, no length scale is forced onto the system.
The origin of the wavelength\revise{, which depends on the excitation frequency, as can be seen on Supplementary Movie SM1,} must thus be dynamical.
\revise{If the forcing is turned off, both perturbations decay quickly, as can be seen on Supplementary Movie SM3,
    confirming that both types of waves are attenuated in the absence of forcing.}

The main purpose of this paper is to explain the features of this instability, the conditions under which it can develop,
and how it saturates.
Let us start by qualitatively describing the phenomenon,
by showing how parametric coupling between longitudinal and transverse waves leads to amplification of perturbations,
and by identifying the resonance condition that selects the pattern.

\subsection{A pattern-forming instability}
Consider the rivulet at rest in the vertical base state.
Let us now impose an additive forcing which is homogeneous in space and harmonic in time.
If the forcing exceeds a certain threshold, which depends on its frequency,
a pattern starts to appear.

The speakers are driven in an antisymmetric configuration, so when one
pushes its diaphragm out, the other pulls it in.
This creates a pressure difference between both sides of the rivulet,
displacing the liquid in the transverse direction.
The rivulet thus behaves as a membrane,
separating the cell into two airtight compartments.
Note that since the speakers are not of infinite size,
the forcing is not perfectly homogeneous:
it is maximal on the midline joining the two speakers and slowly decreases away from it.
All the measurement presented here are made on a \SI{10}{\centi\meter} zone around this midline,
where we checked that the forcing inhomogeneities never exceed \SI{10}{\percent}.

\begin{figure}
    \centerline{\includegraphics[scale=1]{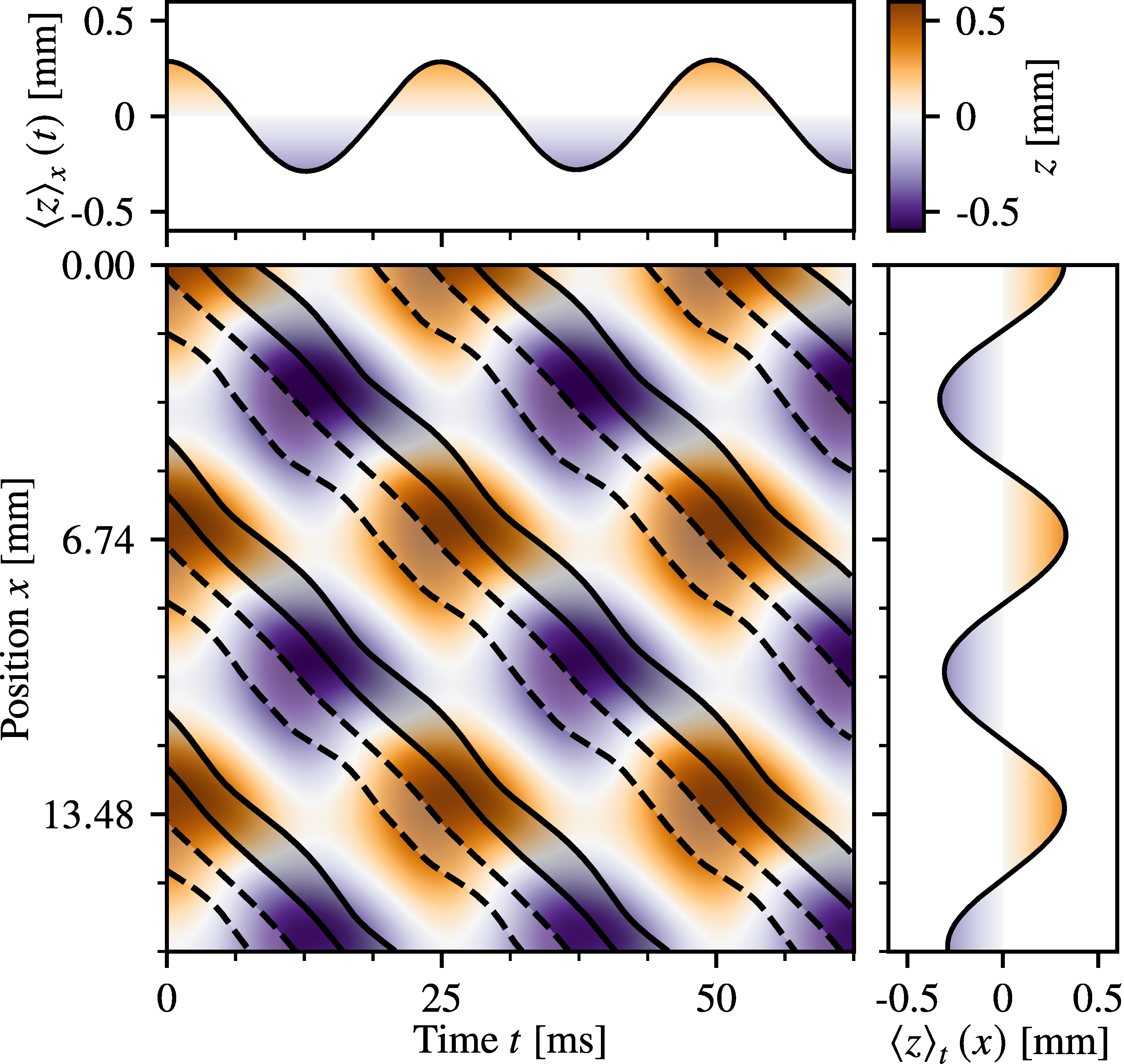}}
    \caption{Spatio-temporal representation of the experimental position $z(x,t)$ of the rivulet as a function of time and space.
    Cell gap $b=\SI{0.58 \pm 0.02}{\milli\metre}$, flow rate $Q=\SI{25.6 \pm 0.9}{\cubic\milli\metre\per\second}$, excitation frequency $\omega_0/(2\pi)=\SI{40}{\hertz}$.
    \\(bottom left): Position $z$ of the rivulet (color scale) as a function of time $t$ and position $x$.
    Darker regions delimited by plain lines correspond to the parts where the rivulet is the heaviest,
        while lighter regions delimited by dashed lines correspond to parts where the rivulet is the thinnest
        (see fig. \ref{fig:spatio_w}).
        \\(right): Time-averaged position \revise{(over \SI{60}{\milli\second})} of the rivulet as a function of space.
        Labeled ticks are spaced by $\lambda=2\pi/k = \SI{6.74}{\milli\meter}$.
        \\(top): Spaced-averaged position  \revise{(over \SI{17}{\milli\meter})} of the rivulet as a function of time.
        Labeled ticks are spaced by $2\pi/\omega_0 = \SI{25.0}{\milli\second}$.
    }
    \label{fig:spatio_z}
\end{figure}

The forcing of the rivulet by the pressure difference between its two
sides is represented mathematically by the rightmost term in equation~\eqref{eq:nsvectoriel}.
This forcing is additive, meaning it induces a linear response from the rivulet.
Thus, since the forcing oscillates harmonically in time at angular frequency $\omega_0$,
the rivulet adopts a movement that is homogeneous in space, and harmonic in time at angular frequency $\omega_0$.
This linear response is always visible on the experimental signal.
Below the instability threshold, it is the only measurable evolution of $z(x, t)$,
but even when the instability develops, this response can still be seen by looking at $\ev{z}_x(t)$,
the space-averaged value of $z$ (see figure~\ref{fig:spatio_z}(top)).

\begin{figure}
    \centerline{\includegraphics[scale=1]{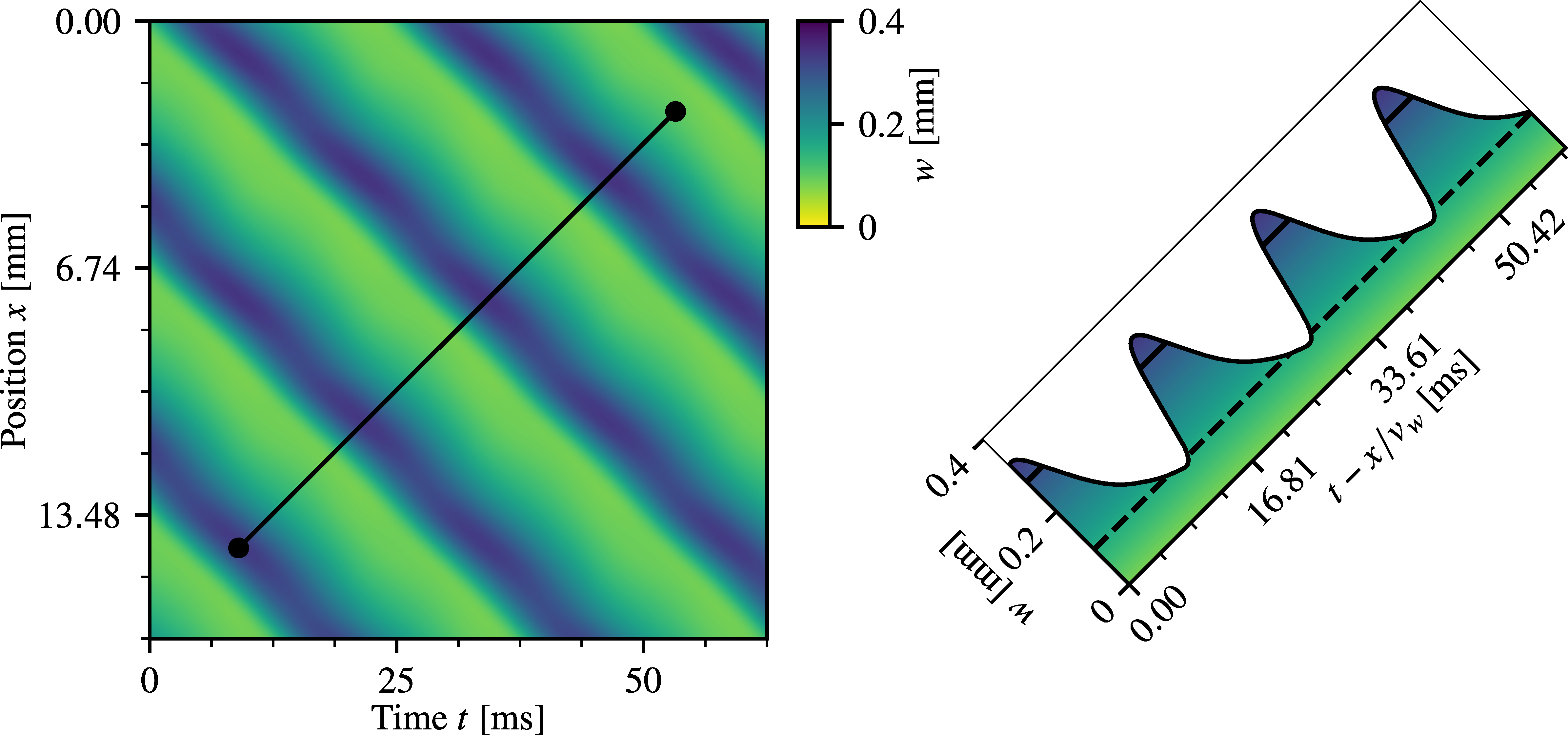}}
    \caption{Spatio-temporal representation of the experimental width $w(x,t)$ of the rivulet as a function of time and space.
    Cell gap $b=\SI{0.58 \pm 0.02}{\milli\metre}$, flow rate $Q=\SI{25.6 \pm 0.9}{\cubic\milli\metre\per\second}$, excitation frequency $\omega_0/(2\pi)=\SI{40}{\hertz}$.
    \\(left): Width $w$ of the rivulet (color scale) as a function of time $t$ and position $x$.
    \\(right): Width of the rivulet interpolated along the plain black line represented on the left plot.
    The plain and dashed lines correspond to the width delimiting the darker and lighter regions on figure \ref{fig:spatio_z},
        respectively.
        The abscissa corresponds to a counter-advected position $x - v_w t$,
        with $v_w=\SI{0}{\milli\meter\per\second}$ the phase speed of longitudinal waves.
        Labeled ticks are spaced by $\sqrt{\lambda^2 + (v_w T_0)^2}$.
    }
    \label{fig:spatio_w}
\end{figure}

Above the instability threshold,
the rivulet adopts a pattern composed of both transverse and longitudinal perturbations.
As can be seen on figure~\ref{fig:spatio_z}~(right),
in addition to the time-dependent response to forcing the rivulet exhibits space-dependent transverse oscillations.
The spatiotemporal transverse signal $z(x,t)$ is a combination of these two contributions.
The rivulet also shows longitudinal oscillations,
which as can be seen on figure~\ref{fig:spatio_w} are a function of $x - v_w t$,
where $v_w = \omega_w^+ / k_w = u_0 + \vcap w_0 k_w$ is the phase speed of longitudinal waves predicted by the dispersion relation~\eqref
{eq:dispersionrelation_w}.
The observation of this pattern bears several questions, which motivate the present study.
\revise{Let us highlight here some surprising key features of the experimental pattern we report,
that will be explained theoretically in the following.}
The first and most obvious feature of the pattern is that,
as can be seen for example on figure~\ref{fig:spatio_z},
both $z$ and $w$ share the same spatial periodicity.
While this indicates that the two waves are probably coupled,
the nature of such a coupling is not clear \textsl{a priori}.
Second, the phase between the transverse and longitudinal patterns is deterministic
(figure~\ref{fig:spatio_z}, bottom left),
which calls for an explanation.
Third, the saturation amplitude of the patterns in $z$ and $w$ is unexplained,
as is the ratio between these different perturbation --- rephrased:
why are the transverse variations of greater amplitude than longitudinal changes ?
\subsection{Mathematical formulation}
\label{subsec:mathematical-formulation}
In order to understand the instability and the pattern it generates,
we write a nonlinear development of equations of equations~\eqref{eq:nsvectoriel} and~\eqref{eq:massconsvectoriel},
following a multiple scale approach~\citep{nayfeh2008, bongarzone2022}.
We will be using a slow timescale $\Tdim = t /\eps^2$,
with $\eps \ll 1$ the small parameter introduced in the previous section.
We also consider the case of weak forcing,
that is to say we take the pressure difference imposed by the speakers $\Pi(t)$ to be of order $\eps^2$.

At order $\eps$, the equations read
\begin{subequations} \label{eq:firstorder}
\begin{align}
    \Lz z_1 &= 0 \\
    \Lw w_1 &= 0
\end{align}
\end{subequations}
i.e. at leading order $z(x,t) = \eps \Zampdim e^{i(\omega_z t - k_z x)} + \cc$ and $w(x,t) = \eps \Wampdim e^{i(\omega_w t - k_w x)} + \cc$,
where $\cc$ stands for \emph{complex conjugate},
and with $(\omega_z, k_z)$ satisfying the transverse waves dispersion relation~\eqref{eq:dispersionrelation_z},
and $(\omega_w, k_w)$ satisfying the longitudinal waves dispersion relation~\eqref{eq:dispersionrelation_w}.

At second order in $\eps$, the equations read:
\begin{subequations} \label{eq:secondorder}
    \begin{align}
    w_0 \Lz z_2 =& \NLzz (z_1, w_1, u_1)
    + \Pi(t) \label{eq:secondorder_z}
    \\
    \Lw w_2 =&\NLww (z_1, w_1, u_1) \label{eq:secondorder_w}
    \end{align}
  \end{subequations}
  where $\NLzz (z_1, w_1, u_1)$ and $\NLww (z_1, w_1, u_1)$ are quadratic nonlinear functions, the expression of which is given in annex~\ref{subsec:annex:order2} ;
and $\Pi(t) = \tilde{\Pi} e^{i\omega_0 t} + \cc$ is the externally imposed pressure
oscillation, homogeneous in space and harmonic in time, of amplitude $\tilde{\Pi}$.
The nonlinear terms $\NLzz (z_1, w_1, u_1)$ and $\NLww (z_1, w_1, u_1)$ contain no contributions able to resonate with $z_1$ and $w_1$.
This can be seen by looking directly at the expressions or by considering symmetry arguments.
The equations \eqref{eq:secondorder} thus reduce to
\begin{subequations}
\begin{align}
    -w_0 {\omega_0}^2 z_2 &= \tilde{\Pi} e^{i\omega_0 t} + \cc \\
    w_2 &= 0
\end{align}
\end{subequations}
the forcing only adds an additive contribution to the transverse pattern:
\begin{align}
    \nonumber
    z(x,t) = \eps \Zampdim e^{i(\omega_z t - k_z x)} + \eps^2 \Fampdim e^{ i(\omega_0 t - 0x)}+ \cc \qqtext{and} w(x,t) = \eps \Wampdim e^{i(\omega_w t - k_w x)} + \cc
\end{align}
The term proportional to $\eps^2 \Fampdim$ on the expression of $z(x,t)$ corresponds to the linear response to the forcing.
The expression of the operator $\Lz$ allows us to write an explicit expression for $\Fampdim$, defined as
$\Fampdim \defeq - \tilde{\Pi} / (w_0\,{\omega_0}^2)$.

At third order in $\eps$,
elimination of the resonant terms leads to the following amplitude equations
\begin{subequations}\label{eq:thirdorder}
\begin{align}
    2\partial_\Tdim \Zampdim
    =&
    -\mu  \qty( 1 + \varepsilon_z \frac{\mucl}{w_0 \mu} \qty(1 + \varepsilon_z )) \Zampdim - i \varepsilon_z \frac{{\omega_0}^2}{w_0 u_0 k_w} \conjugate{\Fampdim} \Wampdim \nonumber
    \\
    & + 7 i {k}^3 u_0 \abs{\Zampdim}^2 \Zampdim + i  {k}^3 u_0 \qty(- \varepsilon_z \frac54 + \frac{4 \varepsilon_w}{kw_0}) \abs{\Wampdim}^2 \Zampdim
    \label{eq:order3_Z}
    \\
    2\partial_\Tdim \Wampdim
    =&
    - \mu \Wampdim - i  \varepsilon_w k_w \omega_0 \qty( \frac{{\omega_0}}{k_w \vcap} -\varepsilon_w w_0 k_z )  \Fampdim \Zampdim \nonumber
    \\
    & -2 i \varepsilon_w  w_0 {k_w}^2 \vcap \abs{k_z\Zampdim}^2 \Wampdim + i \qty( k_w\frac{u_0}{{w_0}^2} -\varepsilon_w \frac32 w_0 \vcap
    {k_w}^4 ) \abs{\Wampdim}^2 \Wampdim
    \label{eq:order3_W}
\end{align}
\end{subequations}
where $\conjugate{A}$ denotes the complex conjugates of $A$ and $\varepsilon_{z,w}$ are each to be replaced by $+1$ or $-1$,
depending on the chosen propagation dispersion ($\varepsilon_w = \pm 1$ for $\omega_w^\pm$, and similarly for $\varepsilon_z$).
The main steps allowing the derivation of equations~\eqref{eq:thirdorder} are shown in annex~\ref{subsec:annex:order3}.
For the equations presented in the made text we made the simplifying assumption $\vcap \approx u_0$,
the full expression of $\partial_\Tdim \Zampdim$ and $\partial_\Tdim \Wampdim$ in the general case is presented in annex~\ref{subsec:annex:order3}.
Let us now try to get a physical understanding of these equations.
The right-hand side comprises three kinds of terms:
dissipative linear damping,
nonlinear cross-coupling,
and third-order nonlinearities, in this order.

For longitudinal waves the linear damping is always the same,
while for transverse waves the damping depends on the branch considered.
When $\varepsilon_z = -1$ ($\omega_z^-$ branch) the multiplicative coefficient for the linear term is $-\mu$,
but the $\varepsilon = +1$ ($\omega_z^+$ branch) is subject to a much more important attenuation, with a coefficient $-\mu(1 + 2\mucl/(w_0\,\mu))$.
Indeed, since the rivulet is thin, the dissipation due to transverse menisci displacement dominates the bulk dissipation:
$\mucl \gg w_0 \mu$, meaning that the $\omega_z^+$ branch is strongly damped.
\revise{Let us physically explain this dependence of dissipation with the phase velocity.}

\revise{In general,
since the rivulet is thin,
the dissipation due to the displacement of the transverse menisci dominates the dissipation in the volume:
$\mucl \gg w_0 \mu$,
the main part of the energy loss comes from the ``friction'' of the menisci on the glass plates that delimit the cell.
A wave on the $\omega_z^-$ branch moves at the velocity $u_0 - \vcap$ in the laboratory frame, thus with respect to the plates.
This velocity is very small (compared to $u_0$ or $\vcap$, which are of the same order of magnitude),
so the menisci are almost motionless with respect to the plates and therefore generate very little friction.
On the contrary, if we consider a wave on the $\omega_z^+$ branch,
the rivulet slides on the plates at the velocity $u_0+\vcap$ in the laboratory frame,
and the menisci are displaced rapidly with respect to the plates,
which is energetically costly.
The $\omega_z^+$ branch is therefore strongly damped:}
this is consistent with the fact that transverse waves $\omega_z^+$ are never
observed experimentally.
For the following we will thus only consider the case $\varepsilon_z  = -1$.
The fact that the damping then becomes exactly the same for transverse
and longitudinal waves is a consequence of the simplifying $\vcap \approx u_0$ assumption.

The most interesting parts of equations~\eqref{eq:thirdorder}
are undoubtedly the cross-coupling terms, which imply an interaction between transverse and longitudinal waves.
The sign of each nonlinear coupling term is critical for the amplification mechanism,
and it can only be obtained through a careful derivation of the third-order equations.
\revise{As we will show rigorously in section~\ref{subsec:threshold}, }the interaction of transverse and longitudinal waves is only constructive if the product of both coupling terms is of positive sign,
in the opposite case the interaction only leads to increased damping for both waves.
In order to observe an instability, we must thus have $(-i \varepsilon_z)(- i \varepsilon_w) = +1$
(assuming $\omega_0 \gg \vcap w_0 k^2$, which is always verified in the accessible experimental range).
Since $\varepsilon_z = -1$, this forces $\varepsilon_w = +1$ to be the branch amplified by the instability.

Looking at figure~\ref{fig:spatio_z},
one can confirm that the prediction $(\varepsilon_z, \varepsilon_w) = (-1, +1)$ is well verified experimentally:
we can see that $\omega_w > \omega_z$, with $v_z = \omega_z/k_z \approx 0$.

Note that the form of the coupling term is a signature that the instability is indeed parametric,
as the forcing response $\Fampdim$ acts as a multiplicative parameter coupling the two waves together.
The fact that the linear response to an additive forcing intervenes multiplicatively in the amplitude equations
participates in the instability originality.
This coupling between two waves mediated by a forcing is analogous to the cross-coupling
between progressive and regressive waves in the Faraday instability in an annulus~\citep{douady1989}.

The rightmost terms of the equations~\eqref{eq:thirdorder}
are nonlinear detuning terms.
Indeed in the absence of nonlinear damping, these terms are associated with a purely imaginary prefactor,
translating the fact that as the amplitudes of the waves grow, their frequencies shift accordingly.
These detuning terms are often responsible for the nonlinear saturation of parametric instabilities,
because they shift the oscillation frequency away from resonance.

To conclude this section, note that the general form of equations~\eqref{eq:thirdorder}
could have been guessed by using only symmetry arguments.
If one assume that the evolution equations for $\Zampdim$ and $\Wampdim$
contain only third-order terms proportional to $\Zampdim$, $\Wampdim$, $\Fampdim$ and their complex conjugates,
then necessarily such terms must be compatible with the invariances and symmetries of the problem.

Let us consider the time translation invariance of the problem,
i.e.\ the fact that the equations should not change when the transformation $t\rightarrow t + t_0$ is applied.
This transformation corresponds, in terms of our variables, to the phase shift $\Zampdim, \Wampdim, \Fampdim \rightarrow \Zampdim e^{i\phi}, \Wampdim e^{i\phi + i\psi}, \Fampdim e^{i\psi}$
with $\phi = \omega_z t_0$ and $\psi = \omega_0 t_0$.
The third-order evolution equations, in order to be invariant under this transformation, must then have the following form:
\begin{subequations}
\begin{align}
    \partial_{\Tdim}\Zampdim = a \mu \Zampdim + b  \conjugate{\Fampdim} \Wampdim + c \Zampdim^2 \conjugate{\Zampdim} + d \Wampdim \conjugate{\Wampdim}\Zampdim
    \\
    \partial_{\Tdim}\Wampdim = a' \mu \Wampdim + b' \Fampdim \Zampdim + c' \Zampdim\conjugate{\Zampdim}\Wampdim + d' \Wampdim^2 \conjugate{\Wampdim}
\end{align}
\end{subequations}
which is indeed the same form as equations~\eqref{eq:thirdorder}.

Finally, the space-time reversal symmetry $x, t\rightarrow -x, -t$,
imposes the phase (the real or imaginary character) of the prefactors $a,b, c, d, a', b', c'$ and $d'$,
since it corresponds to the transformation $\Zampdim, \Wampdim, \Fampdim \rightarrow \conjugate{\Zampdim}, \conjugate{\Wampdim}, \conjugate{\Fampdim}$ and $\mu\rightarrow -\mu$.

\subsection{Resonance condition}
\label{subsec:resonance_condition}

Since equations~\eqref{eq:thirdorder} result from the elimination of secular terms,
one can see that the \revise{cross-coupling }term\revise{s} can only be taken into account if a certain resonance condition is met.
\revise{%
Indeed, for the coupling term $\Fampdim e^{ i(\omega_0 t - 0x)} \Zampdim e^{i(\omega_z t - k_z x)} = \Fampdim \Zampdim e^{i((\omega_z + \omega_0)) t - (k_z+0) x)}$ to interact with the longitudinal wave $\Wampdim e^{i(\omega_w t - k_w x)}$,
both terms must oscillate at the same (spatial and temporal) frequency.
}
The resonance, or frequency-matching condition,
translates to the fact that the interaction between the waves have to be resonant for them to exchange energy and potentially amplify one another.

This behaviour is analogous to a three-wave resonant interaction,
where an algebraic relationship must be satisfied between the waves for the triadic interaction to take place \revise{\citep{simmons1969, martin1972, phillips1981, hammack1993}}.
\revise{It is this mechanism which is responsible, for example, for pattern generation due to nonlinear interaction between gravito-capillary surface waves~\citep{mcgoldrick1965, moisy2012, haudin2016}}.
Here the waves are a transverse wave, a longitudinal wave, and the $\Fampdim e^{i\omega_0 t} + \cc$ contribution to $z(x,t)$ (linear response to forcing),
which can be though of as a 0-wavenumber transverse wave.

The resonance condition thus mathematically reads
\begin{subequations}\label{eq:rescond}
\begin{alignat}{2}
    \omega_w&{}-{}&\omega_z &= \pm \omega_0
    \label{eq:rescond_w}
    \\
    k_w&{}-{}&k_z &=\pm 0
    \label{eq:rescond_k}
\end{alignat}
\end{subequations}
since both temporal and spatial frequencies must match for resonance to take place.
\begin{figure}
    \centerline{\includegraphics[scale=1]{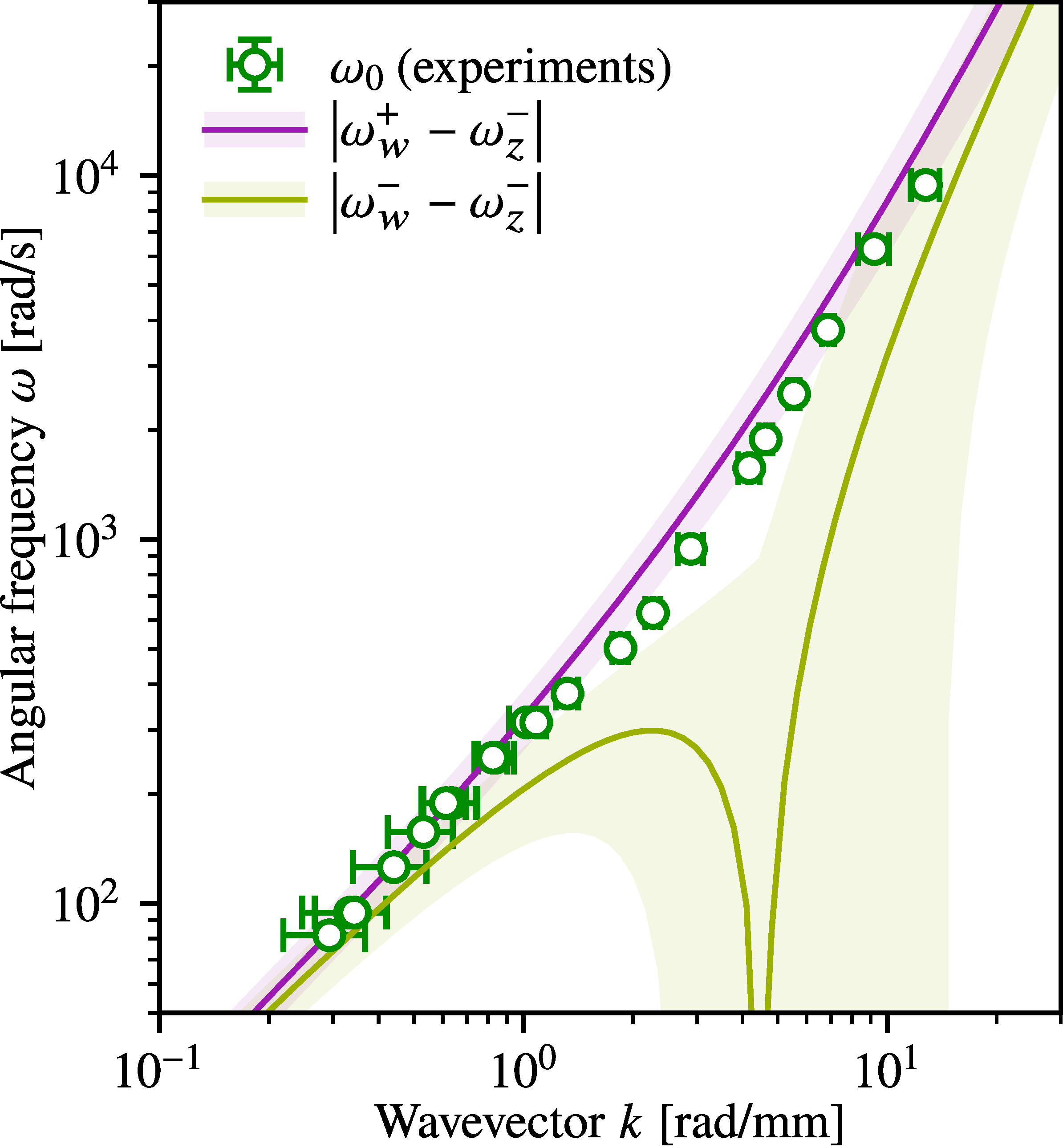}}
    \caption{Relationship between $k$ and $\omega_0$.
    Symbols correspond to experimental measurements of the wavenumber $k$
        (which is the same for transverse and longitudinal waves).
        Experiments done with cell gap $b=\SI{0.6}{\milli\metre}$ and flow rate $Q=\SI{26 \pm 1}{\cubic\milli\metre\per\second}$.
        Lines correspond to $\abs{\omega_w^+ - \omega_z^-} = k\vcap (1 + k w_0)$, i.e. $\varepsilon_z\varepsilon_w = -1$ (purple) and
        $\abs{\omega_w^- - \omega_z^-} = k\vcap \abs{1 - k w_0}$, i.e. $\varepsilon_z\varepsilon_w = +1$ (ocher).
        The line curves are computed without fitting, using the experimental values of the parameters.
        The theoretical prediction corresponds to the modes $\omega_w^+, \omega_z^-$ being unstable (purple curve).
        Note that the experimental points often fall slightly to the right of the curve,
        i.e. the wavevectors $k$ are larger than expected: this is because the points were recorded at a finite amplitude
        for which nonlinear detuning is measurable (see section~\ref{subsec:nonlinear}).
    }
    \label{fig:rescond}
\end{figure}
Equations~\eqref{eq:rescond} impose a necessary condition to the existence of the cross-coupling term, and thus of the instability.
They have straightforward implications for the pattern features.
Equation~\eqref{eq:rescond_k} provides a theoretical explanation for the fact that both transverse and longitudinal waves share the same wavenumber,
and thus the same spatial wavelength.
Equation~\eqref{eq:rescond_w} predicts a deterministic relation between the different frequencies.
The quantity $\omega_w - \omega_z$ can take two values, depending on the value of the product $\varepsilon_-\varepsilon_+$.
The result is plotted on figure~\ref{fig:rescond}.
They show an excellent agreement with our prediction,
that the excited modes are the $\omega_w^+, \omega_z^-$ modes.

Note that the structure of these equations explain simply why it is possible to observe the instability in a very wide range of excitation frequency.
Indeed, the curve $\omega_0 = \omega_w^+ - \omega_z^-$ as a function of $k$ maps $\mathbb{R}$ to itself:
for every excitation frequency $\omega_0$, there exists a wavenumber $k$ for which the resonance condition can be fulfilled.
Experimentally, we could not observe the instability using frequencies below \SI{10}{\hertz},
for in such condition the wavelength becomes comparable to the size over which the forcing is homogeneous.
\subsection{Physical interpretation of the instability mechanism}
\label{subsec:physics}
Let us now try to convey a qualitative understanding of the physical
effects represented by the nonlinear coupling terms of
equations~\eqref{eq:thirdorder} responsible for the destabilization of
the straight rivulet.

One interesting feature of the present instability is that neither of the two types of waves in the final pattern is unstable on its own in the absence of forcing,
nor does the forcing directly amplify any of the perturbation.
While sinuous and varicose instability are often present in fluid
mechanics, for example when studying jets~\citep{eggers2008, MikhaylovWu2020} or sheets~\citep{villermaux2002}),
usually either one mode becomes unstable on its own, or two or more modes become unstable, compete for energy, and the fastest growing mode wins.
In the system we study here, the forcing only acts as a mere intermediary,
as its only role is to couple together two different type of waves which would not \revisecross{linearly} interact in the absence of forcing.
The two coupling terms in equations~\eqref{eq:order3_Z} and~\eqref{eq:order3_W} are mathematical representation of the two ways that longitudinal and transverse waves can influence one another.

\begin{figure}
    \centerline{\includegraphics[scale=1]{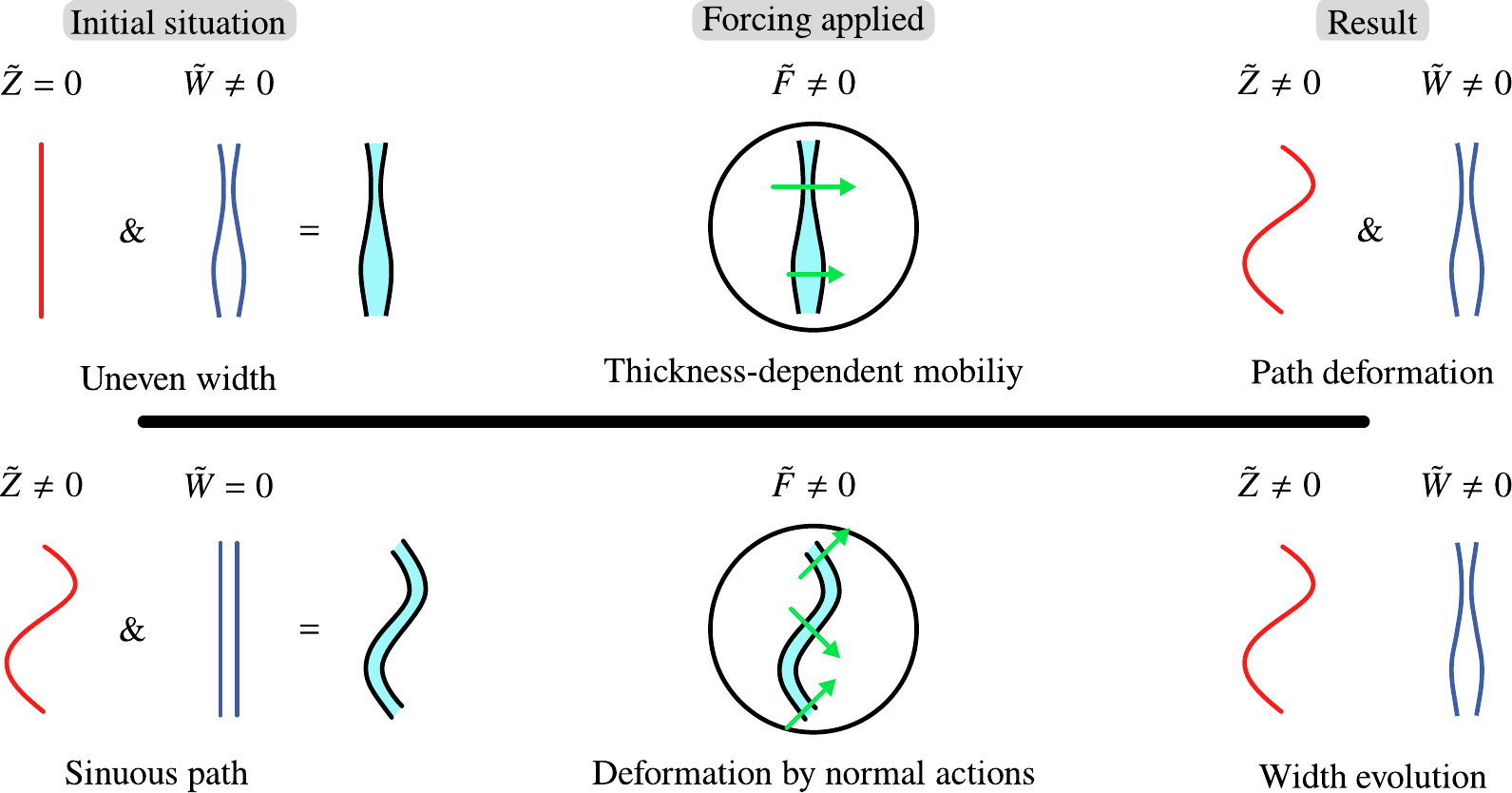}}
    \caption{Schematic illustration of the crossed-amplification mechanism.
    The length of the arrows represent the intensity of the displacement locally imposed to the rivulet.\\
        (top) Deformation of the fluid path due to space-varying mass repartition on the rivulet.\\
        (bottom) Change in the width profile due to the rivulet path being non-straight.\\
    Note that as the forces are drawn, the sinuose and varicose deformation enhance each other.
    If the directions of the arrows are reversed but the sinuosity stays the same
        (corresponding to a change of sign in force while staying in the reference frame moving with the sinuosity, i.e. to transformation~\eqref{eq:explanation_transform1}),
    then the interaction only stays constructive if the width modulation is moved half a wvelength downwards, i.e.~\eqref{eq:explanation_transform2} must be verified. \rerevise{(See also Supplementary figure SF1.}
    }
    \label{fig:mechanism_drawing}
\end{figure}

The $\conjugate{\Fampdim}\Wampdim$ term in equation~\eqref{eq:order3_Z} that governs the evolution of the path $\Zampdim$ translates the fact that
a rivulet with uneven width will see its path deformed when displaced transversally under the action of the forcing,
as illustrated on figure~\ref{fig:mechanism_drawing} (top).
Indeed, if a rivulet presenting width modulations is moved transversally, then the thicker parts of the rivulet, having more inertia,
will be less mobile than the thinner parts of the rivulets, which carry a lesser mass of liquid.
Since different parts of the rivulet, having different mass, will displace transversally along different lengths,
this will lead to a deformation of the path followed by the rivulet.
Following this reasoning, it is straightforward that the spatial periodicity of the path deformation will be the same as the width modulation periodicity.
This allows for a qualitative understanding of condition~\eqref{eq:rescond_k} (the two waves share the same wavelength).

The $\Fampdim\Zampdim$ term in equation~\eqref{eq:order3_W} that governs the evolution of the width perturbation $\Wampdim$ represent the fact that
a rivulet with a curved path will see its width profile evolve when displaced under the action of the forcing,
as illustrated on figure~\ref{fig:mechanism_drawing} (bottom).
Let us considering a rivulet presenting a period curved path, being moved like a membrane under the action of differential pressure between both sides of the rivulet.
At places where the curvature is of the same sign as the pressure displacement force,
the fluid will accumulate, being drawn in from the sides, and the rivulet will get thicker.
On the contrary, places where the curvature is of opposite sign as the pressure force will be depleted in fluid and will be getting thinner,
for the same reason the membrane of an expanding rubber balloon thins.
As before, following this reasoning we also understand that the spatial periodicity of the width modulation will reflect that of the path curvature, validating once more condition~\eqref{eq:rescond_k}.

Finally, Let us considers the same mechanisms that were just exposed in the two preceding paragraphs, but half a forcing period later\rerevise{, as is shown on the bottom part of Supplementary Figure SF1}.
This corresponds to changing the sign of the pressure force / transverse displacement:
formally, this is associated with the
\begin{align}
    \label{eq:explanation_transform1}
    t\rightarrow t+ \pi/\omega_0 && \tilde{\Pi} \rightarrow -\tilde{\Pi} && \Fampdim \rightarrow -\Fampdim && \Zampdim\rightarrow \Zampdim e^{i\pi\omega_z/\omega_0} && \Wampdim\rightarrow \Wampdim e^{i\pi\omega_w/\omega_0}
\end{align}
transformation.
The varicose amplification mechanism will lead to a width modulation of opposite sign compared to the previous situation (since the pressure drop changed sign).
In order for the total mechanism to work this must be on par with the action of the sinuosity amplification mechanism,
which requires the width modulation pattern to be displaced of half a wavelength compared to the sinuous pattern \rerevise{(See Supplementary figure SF1 for an illustration)}:
formally, this corresponds to the
\begin{align}
    \label{eq:explanation_transform2}
    \Wampdim\rightarrow -\Wampdim e^{i\pi\omega_z/\omega_0} = \Wampdim e^{i\pi(\omega_z \pm \omega_0)/\omega_0}
\end{align}
transformation.
Hence, for the mechanism to work at all times, the system must verify $\omega_w = \omega_z \pm \omega_0$: this is precisely the resonance condition~\eqref{eq:rescond_w}.
Both resonance conditions can thus be understood qualitatively by looking at the physical interpretation of the amplification mechanism:
two different effects combine to form the complete amplification cycle.

One obvious follow-up question following this demonstration would be: what are the relative importance of these two effects?
This question, and others, finds its answer in the next section.

\section{Pattern formation and structure}
\label{sec:pattern}

\subsection{Dimensionless formulation}

In order to make the mathematical computations more concise, we use changes of variables to make the system of equations dimensionless.
The timescale choice is straightforward and corresponds to $2/\mu$, the typical wave damping time in the absence of forcing.
As length scale we choose the quantity $\lengthscale \defeq \mu \sqrt{w_0 \vcap / {\omega_0}^3}$, the interpretation of which will become clear in the next subsection.
Thus, we define
\begin{align}
    T \defeq \mu \Tdim/2 && \Famp \defeq \Fampdim/\lengthscale && \Zamp \defeq \Zampdim/\lengthscale && \Wamp \defeq \Wampdim/\lengthscale
\end{align}
which allows us to write the \revisecross{system} \revise{equations~\eqref{eq:thirdorder}} in a compact dimensionless form :
\begin{subequations}
\begin{align}
    \partial_T Z &= - Z + i \phi \conjugate{F} W  + i \qty(\alpha_{ZZ} \abs{Z}^2+ \alpha_{ZW}\abs{W}^2 )Z
    \label{eq:admin_z}
    \\
    \partial_T W &= -  W - i\frac{1}{\phi}  F Z  + i \qty(\alpha_{WZ} \abs{Z}^2+ \alpha_{WW}\abs{W}^2 )W
    \label{eq:admin_w}
\end{align}
\end{subequations}
where $\phi \defeq \frac{1}{u_0 k}\sqrt{\frac{\omega_0\vcap}{w_0}}$ ;
and the detuning prefactors $\alpha_{XX}$ are defined as follows:
\begin{align*}
    \alpha_{ZZ} &\defeq 7\,\Xi\,(k w_0)^3 &  \alpha_{ZW} &\defeq 4 \Xi\,(k w_0)^2 \qty(1 + \frac{5}{16} (k w_0))
    \\
    \alpha_{WZ} &\defeq 2\,\Xi\,\frac{\vcap}{u_0} (k w_0)^4 &  \alpha_{WW} &\defeq \Xi\,k w_0
\end{align*}
where $\Xi \defeq \frac{\mu u_0\vcap}{{\omega_0}^3{w_0}^2} = \frac{u_0 \lengthscale^2}{\mu {w_0}^3}$ characterizes the detuning strength.
The dimensionless coefficient $\phi$ translates the relative efficacy of the two effects that participate in the co-amplification:
amplification of sinuosity due to uneven mass repartition,
and amplification of varicosity due to path curvature
(see section~\ref{subsec:physics} for further details on the physical interpretation of these terms).
It thus quantifies the asymmetry of the instability mechanism:
thick rivulets moved at low frequency have $\phi \gg 1$ and are unstable because of the path deformation due to width heterogeneity,
while thin rivulets moved at high frequency have $\phi \ll 1$ and are mostly unstable because of curvature induced fluid concentration along the rivulet.
All the experiments presented in this study display $\phi > 1$,
meaning that the main amplification stage in the mechanism is the differential movement of the path due to the spatially inhomogeneous inertia of the rivulet.

\subsection{Instability threshold}
\label{subsec:threshold}
We will now turn to the study of the pattern near the instability threshold:
in all the following,we will assume $\abs{Z} \sim \epss$ and $\abs{W} \sim \epss$, with $\epss\ll 1$.
We define the vector $\ket{U} \defeq \smqty(Z \\ W)$ which condenses all the information about the state our system is in,
using a concise notation.
We borrow the bra-ket notation to our quantum mechanics colleagues as it is particularly useful when dealing with linear algebra problems.
Since $\norm{U} \ll 1$, equations~\eqref{eq:admin_z} and ~\eqref{eq:admin_w} can be simplified to
\begin{align}
    \partial_T \ket{U} &= \evmat \ket{U} \qqtext{with} \evmat=\mqty(-1 & i\phi \conjugate{F} \\ -i \frac1\phi F & -1)
\end{align}
being a linear evolution operator.

\begin{figure}
    \centerline{\includegraphics[scale=1]{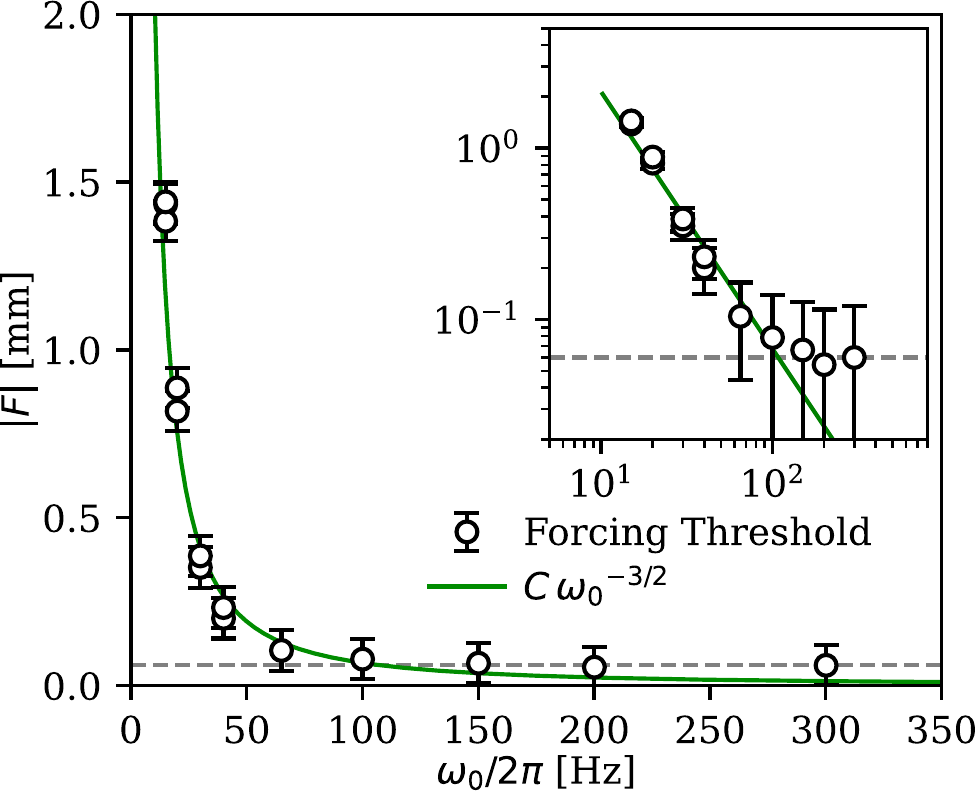}}
    \caption{Threshold transverse displacement as a function of excitation frequency.
    Experiments done with cell gap $b=\SI{0.6}{\milli\metre}$ and flow rate $Q=\SI{26 \pm 1}{\cubic\milli\metre\per\second}$.
    The black dots corrrespond to experimental measurements.
    The dashed grey line represents the resolution limit of our measurement method,
    it corresponds to 1/5th of the pixel spacing.
    The full green line is a one-parameter fit corresponding to a power law of exponent $-3/2$, the multiplicative factor being free.
    Inset: same data on a log--log scale.
    }
    \label{fig:threshold}
\end{figure}

The trace of the matrix $\evmat$ is $\tr (\evmat) = -2 < 0$.
Since it is the sum of the eigenvalues $\evmat$,
we can conclude that there always exists at least one mode that is linearly damped.
The determinant reads $\Delta (\evmat) = 1 - \abs{F}^2$, it is the product of the eigenvalues of $\evmat$.
The instability develops if and only if one mode is linearly amplified,
i.e.\ if one of the eigenvalues of $\evmat$ has a positive real part.
This is only the case if $\Delta (\evmat)$ is negative,
which happens when the forcing exceeds a critical threshold equal to one in dimensionless units.
The instability thus develops if and only if the condition
\begin{align}
    \abs{F}^2 > \abs{\Famp_c}^2 \defeq 1 \quad
    \Leftrightarrow
    \quad
    \abs{\Fampdim} > \abs{\Fampdim_c} \defeq \mu \sqrt{\frac{w_0 \vcap }{ {\omega_0}^3}} = \lengthscale
    \label{eq:forcingthreshold}
\end{align}
is fulfilled.
The interpretation of $\lengthscale$ is now clear:
it corresponds to the minium transverse displacement needed to develop the instability.
This provides us with a scaling law for the threshold forcing amplitude: $\abs{\Fampdim_c} \propto (\omega_0)^{-3/2}$.
This scaling law is shown on figure~\ref{fig:threshold} to be verified by experimentally.
Interestingly,
the prefactor value obtained by computing the best fit (\SI{33.7}{\milli\meter\times \hertz^{-3/2}})
\revise{is the same order of magnitude but} does not correspond \revise{exactly} to the theoretical prefactor computed using the analytical formula~\eqref{eq:forcingthreshold} (\SI[separate-uncertainty=true]{16.4 \pm 3,3}{\milli\meter\times \hertz^{-3/2}}).
\revise{Several factors may explain this discrepancy.}
\revise{On the experimental side,} the precise measurement of $\abs{\Fampdim_c}$ itself is difficult,
especially at high frequencies.
\revise{On the theoretical side, this difference may be explained, partially or totally,
    by several factors: the $\eps \ll 1$ approximation, which is not strickly respected in the developped pattern,
the weakly nonlinear method we use, if the nonlinearity is more important than anticipated, or the $u_0\approx \vcap$ approximation, which is not exact.}
The fact that the best fit value is almost exactly a factor 2 off the theoretical result could also indicate an algebraic error in our computations,
even though we could not catch it after re-deriving our results several times.
\subsection{Pattern structure}
After identifying the forcing threshold over which the instability develops, we now turn to the study of the pattern itself.
We ought to understand the structure of the most unstable mode that grows, and compare it to the pattern observed in the experiments.

To do so, we place ourselves in the vicinity of the instability threshold,
assuming $F = F_c + \epss^2\delta F$
with $\abs{F_c} = 1$ and $\epss$ the small parameter introduced in the previous subsection.
This allows us to decompose the evolution matrix into
\begin{align}
    \evmat = \evmatlinear + \epss^2 \evmatsmall = \mqty(-1 & i\phi \conjugate{F_c} \\ -i \frac1\phi F_c & -1) + \epss^2 i \mqty(0 & \phi \conjugate{\delta F} \\ - \frac1\phi \delta F & 0).
\end{align}
The matrix $\evmatlinear$ is diagonalizable and has two eigenvalues, $\lambda_0 = 0$ and $\lambda_- = -2$, associated to the two eigenvectors
\begin{align}
    \ket{V_0} = \frac{1}{\sqrt{1 + \abs{F_c}^2/\phi^2}}\mqty(1 \\ -i F_c/\phi) && \qqtext{and} && \ket{V_-}  = \frac{1}{\sqrt{1 + \phi^2\abs{F_c}^2}}\mqty(-i \phi \conjugate{F_c} \\ 1)
\end{align}
which are of norm 1, using the standard product $\braket{P}{Q} = P^\dagger\, Q$ where $P^\dagger$ represents the adjoint, or trans-conjugate, of $P$.

The eigenvector $\ket{V_0}$ corresponds to a neutral mode of the operator $\evmatlinear$.
It is the mode that will be amplified and that saturates when nearby the threshold instability.
Thus, we can write, at first order in $\epss$, $\ket{U} = \epss A \ket{V_0}$ with $A$ being the dimensionless amplitude of the pattern.

Looking at $\ket{V_0}$, it is thus possible to obtain information about the structure of this neutral mode.
In particular, if the pattern is given by $\ket{V_0}$, then we can write the following relationship between $Z$ and $W$:
\begin{align}
    W = -i \frac{F_c}{\phi}Z \label{eq:relWZF}
\end{align}
where here the $\phi$ factor translates the fact that since one mechanism is more efficient than the other,
then one of the waves will have greater amplitude.
Indeed, in the experimental pattern (figures~\ref{fig:spatio_z} and~\ref{fig:spatio_w})
we observe that the transverse waves have a greater amplitude than the longitudinal waves,
which is coherent with the fact that the $\phi$ factor is always greater than one in the experiments we show in this article.
\begin{figure}
    \centerline{\includegraphics[scale=1]{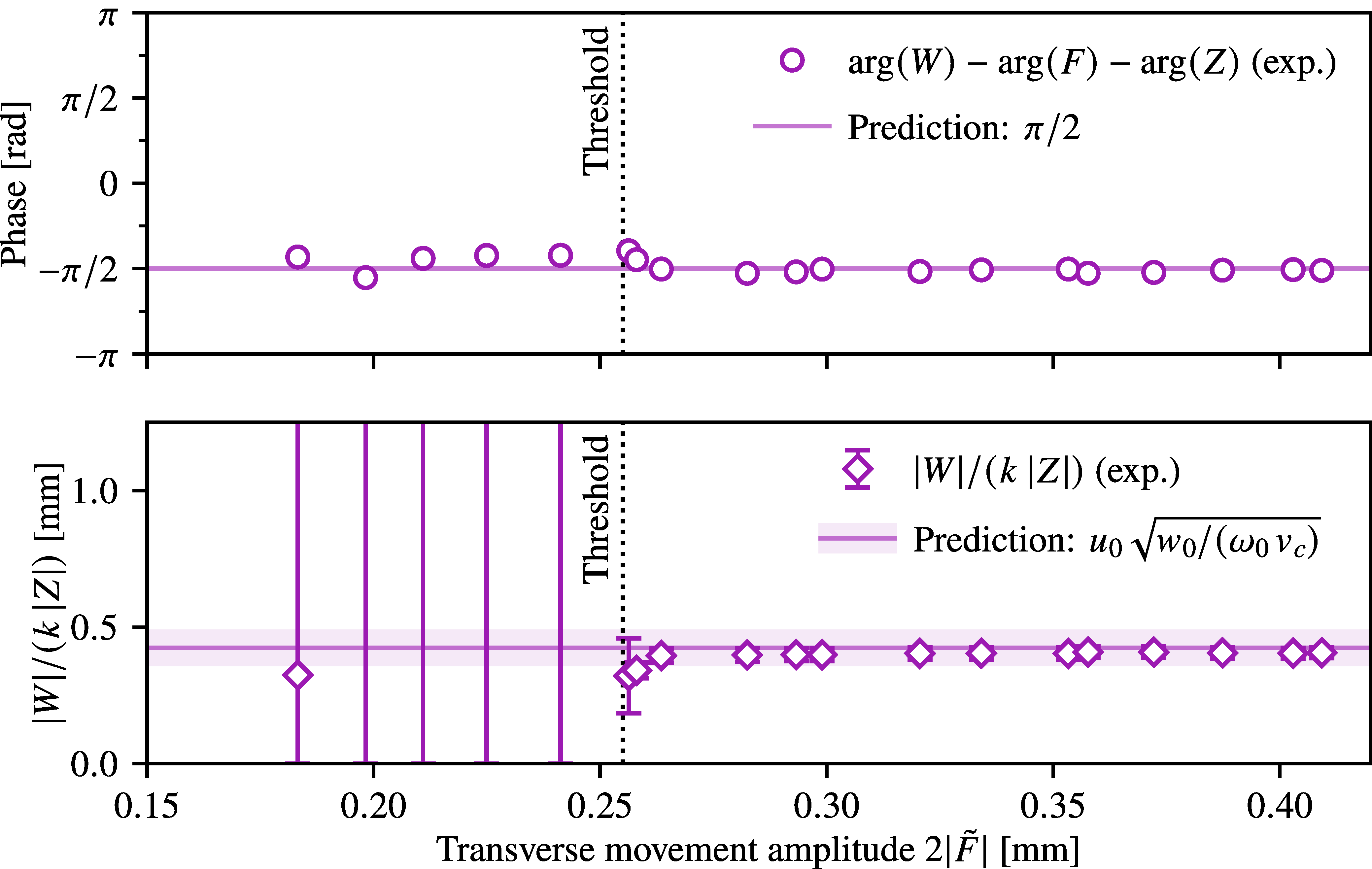}}%
    \caption{Structure of the combined mode.
    The dots always correspond to experimental measurement.
    Cell gap $b=\SI{0.58 \pm 0.02}{\milli\metre}$, flow rate $Q=\SI{25.6 \pm 0.9}{\cubic\milli\metre\per\second}$, excitation frequency $\omega_0/(2\pi)=\SI{40}{\hertz}$.
    \\(top): Relative phase of the waves as a function of the trtansverse movement amplitude.
    There are measurement points below the threshold where we can still observe a weak pattern (see discussion on section~\ref
    {subsec:nonlinear}).
    \\(bottom): Link between the relative amplitude of transverse and longitudinal waves as a function of the transverse movement amplitude.
    The points below the threshold are associated to high experimental uncertainty, since the associated amplitudes are very small (see figure~\ref{fig:saturation}).
    }
    \label{fig:modestructure}
\end{figure}
Equation~\eqref{eq:relWZF} contains all the information about the structure of the instability pattern.
Taking the argument and the norm, we can write
\begin{subequations}
\begin{align}
    \arg W &= \arg{Z} + \arg{F} - \frac{\pi}{2}
    \label{eq:argrelWZF}
    \\
    \abs{W} &= \abs{Z}/\phi = k \abs{Z}\  u_0\sqrt{w_0 / (\omega_0\,\vcap)}
    \label{eq:absrelWZF}
\end{align}
\end{subequations}
Equation~\eqref{eq:argrelWZF} predicts the relative phase between the longitudinal wave, the transverse wave and the response to forcing.
The arguments of $Z$, $F$ and $W$ correspond to three degrees of freedom of the system.
Since any pair of these can be arbitrarily changed by redefining the origins of space (by the transformation $x \rightarrow x+ x_0$)
and time (by the transformation $t \rightarrow t+ t_0$), this relation completely constrains the system.
As can be seen on figure~\ref{fig:modestructure} (top), the relation~\eqref{eq:argrelWZF} is verified experimentally with excellent precision.
Note that signature of this phase relation can be seen when observing directly the pattern.
For example, on figure~\ref{fig:spatio_z},
extrema of $w$ (centre of dark or light shaded zones in the central plot, corresponding to $\arg(W) \equiv 0 [\pi]$)
are never found in places where $z$ is extremal (i.e.\ where $\arg(Z) + \arg(F)  \equiv 0 [\pi]$).
On the same figure, by looking carefully at the intersections between extrema of $w$, extrema and zeros of $Z$ (top plot), and extrema and zeros of $F$ (right plot),
one can recover~\eqref{eq:argrelWZF}.

Complementary to~\eqref{eq:argrelWZF},
relationship~\eqref{eq:absrelWZF} predicts the relative amplitude between the two waves forming the pattern.
The norms of $Z$, $F$ and $W$ correspond to three degrees of freedom of the system.
The amplitude of $F$ is directly imposed by the forcing,
and the pattern amplitude $\abs{A}$ depends on non-linear saturation effects,
hence equation~\eqref{eq:absrelWZF} closes the system and determines completely the pattern.
As one can see on figure~\ref{fig:modestructure} (bottom),
this relationship is very well recovered by the experiments.

In order to completely characterize the instability,
we still need to understand the saturation of the pattern amplitude $A$, which is the object of the next subsection.

\subsection{Nonlinear detuning and saturation}
\label{subsec:nonlinear}
We now turn to finer effects due to nonlinear terms: frequency detuning and amplitude saturation.
To do so, we allow the amplitude $A$ to vary with a slow dimensionless timescale $\tau  = T / \epss^2$, so that equations~\eqref{eq:admin_z} and ~\eqref{eq:admin_w} can be written,
up to the third order in $\epss$, in the form
\begin{align}
    \partial_{\tau}\ket{U} = (\evmatlinear + \epss^2 \evmatsmall)\ket{U}+ i \mqty(\alpha_Z & 0 \\ 0 & \alpha_W) \epss^2\abs{A}^{2} \ket{U}
    \label{eq:nonlinearU}
\end{align}
where we defined $\alpha_z$ so that
\begin{align*}
    \epss^2\abs{A}^2 \alpha_Z = \alpha_{ZZ}\abs{Z}^2 + \alpha_{ZW}\abs{W}^2 = {\epss^2 \abs{A}^2} \frac{\alpha_{ZZ} + \alpha_{ZW} \abs{F_c}^2  /\phi^2}{1 + \abs{F_c}^2 /\phi^2}
\end{align*}
and similarly so for $\alpha_W$.
We also have to expand $\ket{U}$ in powers of $\epss$: $\ket{U} = \epss \ket{U_1} + \epss^2\ket{U_2} + \epss^3\ket{U_3}+ \dots$
Equation~\eqref{eq:nonlinearU} at order 1 yields the familiar result $\ket{U_1} = A(\tau)\ket{V_0}$.
By going at higher order, we are now searching to characterize the evolution of $A$ with $\tau$.
At order 2 the equation leads to $\ket{U_2} = \ket{0}$, while the third order development yields
\begin{align}
    \partial_{\tau}A \ket{V_0} = \evmatlinear\ket{U_3} + A \evmatsmall\ket{V_0}+ i \abs{A}^{2}A\mqty(\alpha_Z & 0 \\ 0 & \alpha_W)  \ket{V_0}
\end{align}
which gains to be rewritten in the form
\begin{align}
    \ket{N} \defeq \evmatlinear\ket{U_3} = \partial_{\tau}A \ket{V_0} - iA  \mqty(0 & \phi \conjugate{\delta F}\\ -\frac1\phi \delta F & 0) \ket{V_0} - i \abs{A}^{2}A\mqty(\alpha_Z & 0 \\ 0 & \alpha_W)  \ket{V_0}\
\end{align}
Using the Fredholm alternative, we find that $\ket{N}$ must be orthogonal to the kernel of $\evmatlinear^\dagger$.
This can be justified using the simple following reasoning:
a weak writing of the equality $\evmatlinear\ket{U_3} = \ket{N}$ is that any vector $\ket{P}$ must verify $\braket{P}{N} = \bra{P} \evmatlinear\ket{U_3} = \bra{U_3}{\evmatlinear}^\dagger\ket{P}$.
In particular, if $\ket{P} \in \ker({\evmatlinear}^\dagger)$, that is to say if ${\evmatlinear}^\dagger\ket{P} = 0$,
then $\braket{P}{N} = 0$, recovering that $\ket{P}$ is orthogonal to $\ket{N}$.

The kernel of ${\evmatlinear}^\dagger$ is the span of the vector $\ket{W_0}$, the eigenvector of ${\evmatlinear}^\dagger$ of norm unity associated to the 0-eigenvalue
\begin{align*}
    \ket{W_0} \defeq \frac{1}{\sqrt{1 + \phi^2{F_c}^2}}\mqty(1 \\ -i \phi F_c)
\end{align*}
which resembles closely $\ket{V_0}$ since $\evmat$ is close to be hermitian.
The condition $\braket{W_0}{N} = 0$ yields the following amplitude equation:
\begin{align}
(1 + \abs{F_c}^2)\partial_{\tau}A = A (\conjugate{F_c} \delta F + {F_c} \conjugate{\delta F}) + i \abs{A}^2 A (\alpha_Z + \alpha_W\abs{F_c}^2 )
\label{eq:amplitudelacking}
\end{align}
This equation presents an amplification term as well as a nonlinear detuning, but no saturation of amplitude.
In contrary of other parametric systems such as the parametric pendulum described by the Mathieu equation,
detuning here does not force a departure from resonance.
This is because although nonlinear detuning shifts the frequencies away from the $F=F_c$ equilibrium values,
the system adapts by modifying the wavelength of the pattern in order to preserve the resonance conditions~\eqref{eq:rescond_k} and~\eqref{eq:rescond_w}.

In order for the amplitude equation to have bounded solution,
it is necessary to add higher-order terms into equation~\eqref{eq:amplitudelacking}.
We can do so without needing to compute the prefactor explicitly,
by considering expressions that respect the symmetries of the problem.
In particular, the symmetry $x\rightarrow x + x_0\Rightarrow A\rightarrow Ae^{i \varphi}$ forces the supplementary term to be of the form $A^n\conjugate{A}^m$ with $n = m+1$.
Adding in equation~\eqref{eq:amplitudelacking} the smallest order term that respects this symmetry of the problem leads to writing the ansatz
\begin{align}
(1 + \abs{F_c}^2)\partial_{\tau}A = A (\conjugate{F_c} \delta F + {F_c} \conjugate{\delta F}) + i \abs{A}^2 A (\alpha_Z + \alpha_W\abs{F_c}^2 ) - \sigma \abs{A}^4 A
\end{align}
where $\sigma$ is a complex number, with positive real part so that the last term is responsible for the saturation.
We now write $A(\tau)$ in the polar form $R(\tau) e^{i \Omega \tau}$ with $R > 0$.
The evolution equations for $R$ and $\Omega$ are then:
\begin{subequations}
\begin{align}
    \partial_\tau R  &=  (F-F_c) R  - \frac{\Re(\sigma)}{2} R^5
    \label{eq:amplitude}
    \\
    \Omega  &=  \frac{\alpha_Z + \alpha_W}{2} R^2 - \frac{\Im(\sigma)}{2} R^4 \approx \frac1{2}({\alpha_Z + \alpha_W}) R^2\quad \text{for $R \ll 1$}
    \label{eq:detuning}
\end{align}
\end{subequations}
where without loss of generality we redefined the origin of times in order to have $\arg{F} = \arg{\delta F} = \arg{F_c} = 0$.
\begin{figure}
    \centerline{\includegraphics[scale=1]{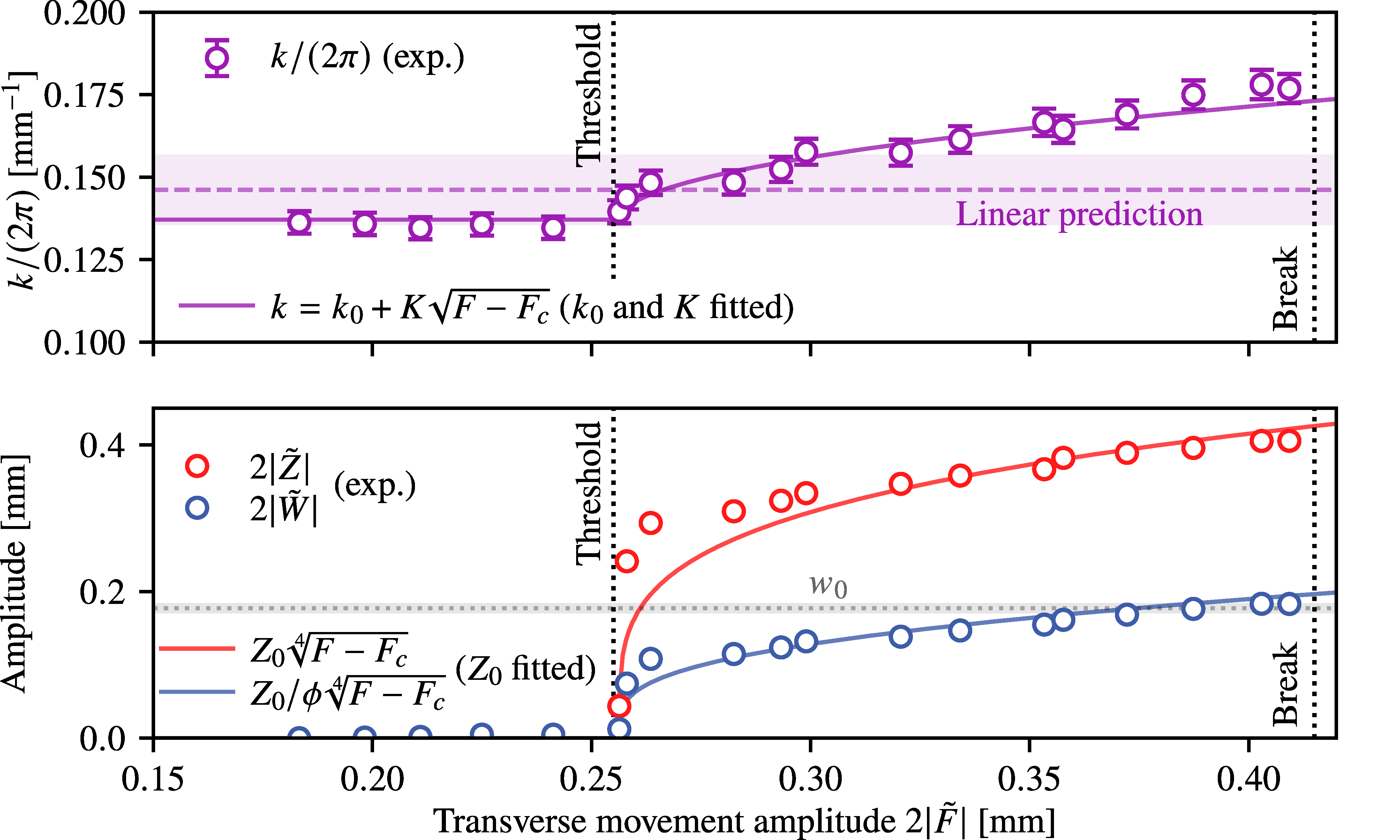}}%
    \caption{Non-linear detuning and saturation.
    The dots always correspond to experimental measurement.
    Cell gap $b=\SI{0.58 \pm 0.02}{\milli\metre}$, flow rate $Q=\SI{25.6 \pm 0.9}{\cubic\milli\metre\per\second}$, excitation frequency $\omega_0/(2\pi)=\SI{40}{\hertz}$.
    \\%
    (top): Spatial detuning as a function of forcing.
    The linear prediction is coherent with the 0-amplitude instability wavelength.
    The solid line corresponds to the function $k \rightarrow k_0 + K\sqrt{F - F_c}$ where the value of $k_0$ and $K$ are the ones that best fit the data.
    \\%
    (bottom): Amplitude evolution of the instability as a function of the forcing response amplitude.
    The filled dots correspond to experimental points.
    Under the threshold ($\Fampdim_c \approx \SI{0.255}{\milli\meter}$), the values of both $\Zampdim$ and $\Wampdim$ are very close to zero.
    The filled lines correspond to the function $\Zampdim, \Wampdim \rightarrow Z_0 \sqrt[4]{F - F_c}, Z_0/\phi \sqrt[4]{F - F_c}$ where the value of $Z_0$ is the one that best fits the data.
    }
    \label{fig:saturation}
\end{figure}
Equation~\eqref{eq:amplitude} predicts that the amplitude of the unstable pattern $\abs{A} = R$ grows when $F > F_c$,
until it reaches a saturation value that is proportional to $\sqrt[4]{F - F_c}$: this is experimentally confirmed on figure~\ref{fig:saturation} (bottom).
Equation~\eqref{eq:detuning} predicts that the frequencies of the waves grow by a quantity $\mu \Omega / 2$.
In order for $\omega_z$ and $\omega_w$ to stay close from their respective dispersion relations,
while still verifying the resonance condition,
the pattern wavelength must also be modified.
In the regime $k \ll 1/w_0$, one can write $\omega_w \approx u_0 k$.
Hence, the wavelength is modified by a quantity $K \approx \Omega / \vcap$.
This predicts a displacement $k - k_0 \propto \sqrt[2]{F - F_c}$ which,
as can be seen in figure~\ref{fig:ft_nonlinear} (top), describes the experiments well.

The imperfect match between our model and experimental data
can be at least partially explained by the fact that the excitation takes place in a finite zone.
Near the threshold, the length over which the instability develops diverges.
Since our system is finite, we thus do not see the fully developed instability in the form that our model rigorously describes.
The match becomes better as the forcing goes further away from the threshold.

\subsection{Rivulet breakup}

As can be seen on figure~\ref{fig:saturation}, the instability stops existing above a certain forcing.
This is because of rivulet breakup:
above a critical movement amplitude, the rivulet breaks at one point into two disconnected top and bottom parts.
\revise{This processus is shown on the Supplementary Movie SM4.}
The bottom part of the rivulet then falls down rapidly at $u_0$,
while the fluid of the top part retracts until forming a large droplet, which then falls down vertically.
Between the two parts, air can communicate between the right and left halves of the cell.
This means that the acoustic forcing becomes extremely inefficient,
since the rivulet no longer separates both halves in an airtight manner: the membrane is breached.
When the top part of the rivulet as fallen down low enough,
the membrane-like property of the rivulet is restored and the forcing is effective again:
the instability grows and break, repeating the process periodically.

The reason the rivulet breaks is that the two side menisci bordering the liquid filament enter in contact with each other.
This happens when the local width of the rivulet reaches 0,
i.e.\ in our model when $2\Wampdim = w_0$.
Indeed, this is clearly visible in figure~\ref{fig:saturation}:
the breakup (indicated by a black dotted line) happens when $2\Wampdim \approx w_0$,
and we never observe width variations of amplitude significantly greater than the rest width.

\begin{figure}
    \centerline{\includegraphics[scale=1]{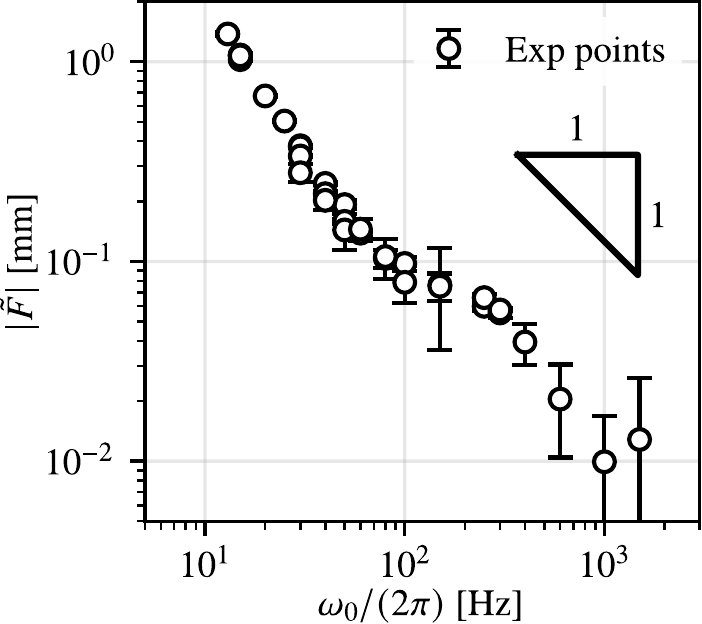}}%
    \caption{Forcing at breakup for varying excitaqtion frequencies.
    Experiments done with cell gap $b=\SI{0.6}{\milli\metre}$ and flow rate $Q=\SI{26 \pm 1}{\cubic\milli\metre\per\second}$.
    The curve was obtained by measuring the linear response to forcing $\Fampdim$ at the maximum forcing amplitude,
        just before breakup.
        It is presented in log--log scale, with a -1 power law as guide for the eyes.
    }
    \label{fig:breakup}
\end{figure}

On figure~\ref{fig:breakup} we show the breakup forcing as a function of the excitation frequency.
The points seem to align along a relation of the type $\Fampdim_\text{break}\propto 1/\omega_0$,
although the physical origin of this scaling law is unclear.

\section{Conclusion}
\label{sec:conclusion}

To end this article,
we first highlight how several key characteristics of the system we study can be retrieved concisely
by representing the experimental signals in the frequency space.
We then present a summary of our study as well as promising directions for future research.

\subsection{Representation in the Fourier space}
\label{subsec:fourier}

\begin{figure}
    \centerline{\includegraphics[scale=1]{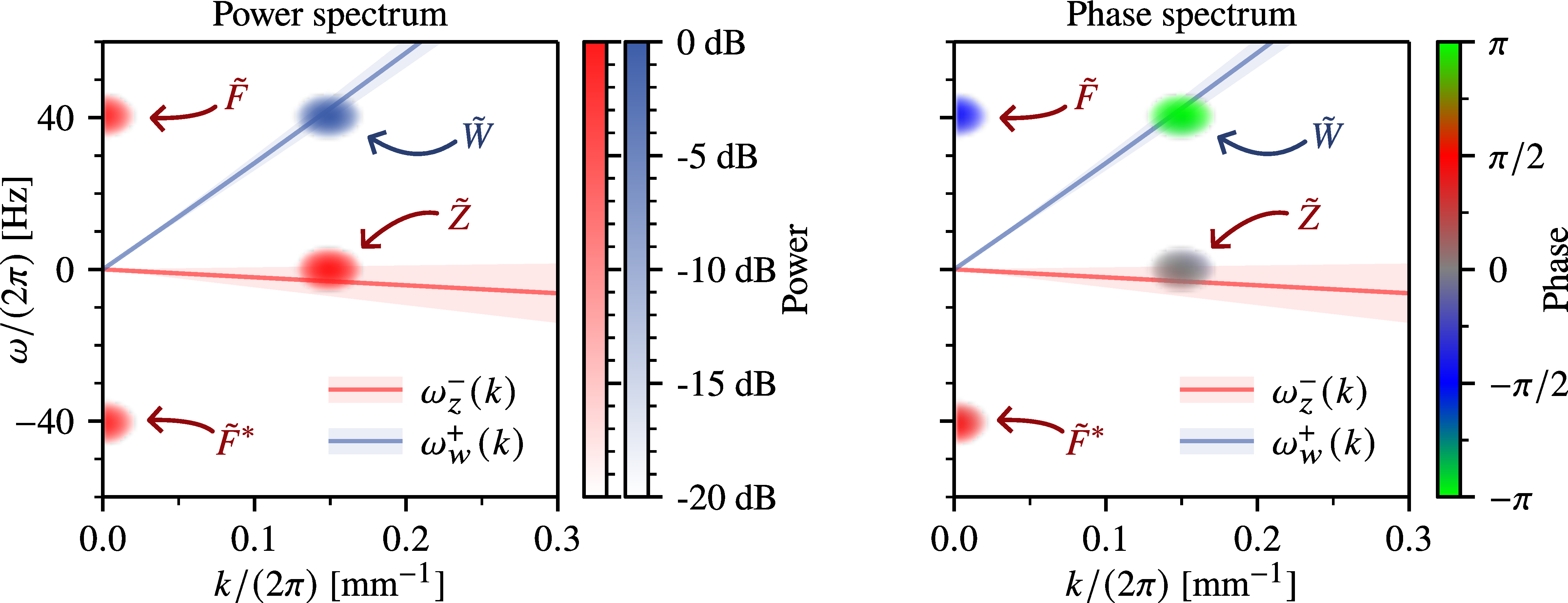}}
    \caption{Experimental rivulet path $z$ and width $w$, represented in the Fourier space.
    Cell gap $b=\SI{0.58 \pm 0.02}{\milli\metre}$, flow rate $Q=\SI{25.6 \pm 0.9}{\cubic\milli\metre\per\second}$, excitation frequency $\omega_0/(2\pi)=\SI{40}{\hertz}$.
    The filled lines represent the dispersion relations of transverse (red) and longitudinal (blue) waves.
    Both graph were obtained using zero-padding of the experimental signal.
    \\ (left) Power spectrum of $\hat{z}(k, \omega)$ (red signal) and $\hat{w}(k, \omega)$ (blue signal)
    For both signals, the color intensity corresponds to scale going from a reference value of \SI{0}{\deci\bel}
    (corresponding to the most intense value of this signal)
    to \SI{-20}{\deci\bel}.
    \\ (right) Phase spectrum of $\hat{z}(k, \omega)$ and $\hat{w}(k, \omega)$.
    Color hue corresponds to the phase of the signal, while color intensity represents the signal strength.
    As for the left plot, the color scale fades to white under 20 dB of the reference signal.
    }
    \label{fig:ft_phase}
\end{figure}

\begin{figure}
    \centerline{\includegraphics[scale=1]{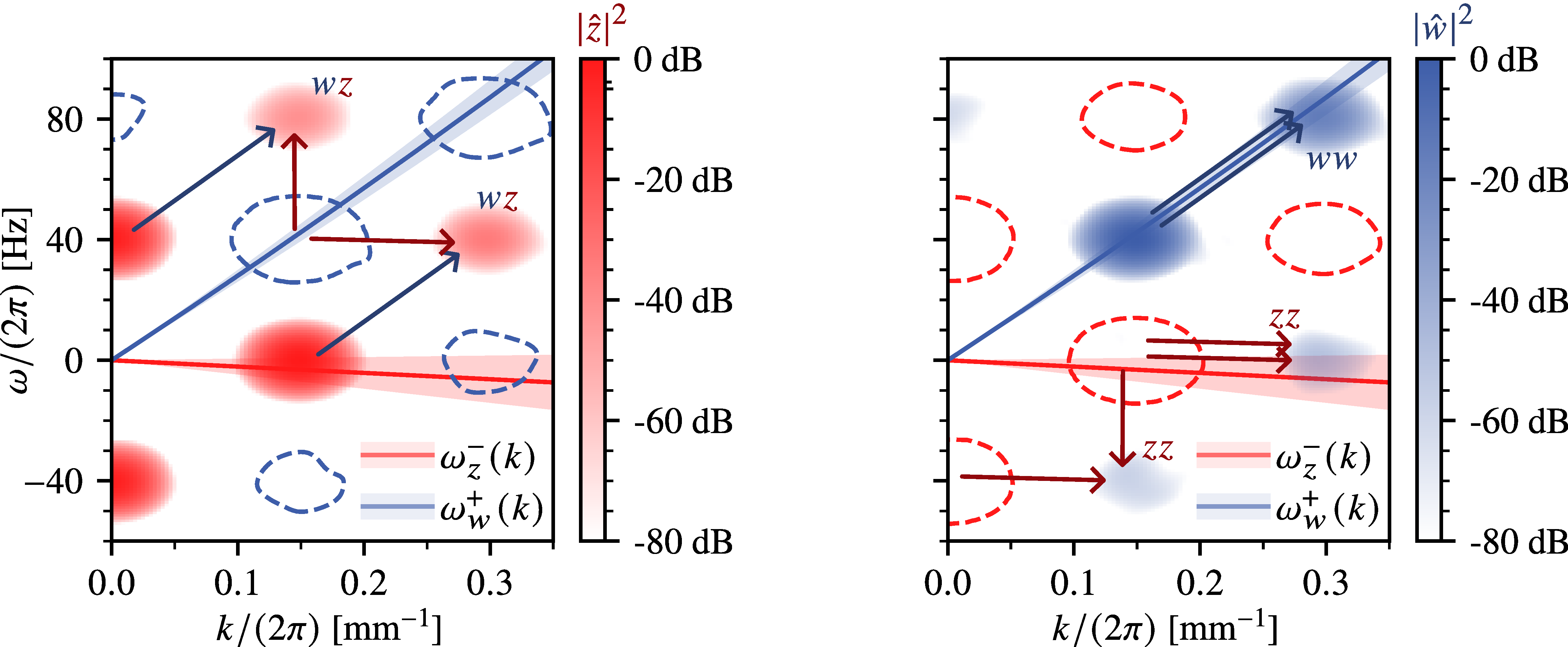}}%
    \caption{Experimental signals of the rivulet path $z$ and width $w$, represented in the Fourier space.
    Cell gap $b=\SI{0.58 \pm 0.02}{\milli\metre}$, flow rate $Q=\SI{25.6 \pm 0.9}{\cubic\milli\metre\per\second}$, excitation frequency $\omega_0/(2\pi)=\SI{40}{\hertz}$.(same conditions as fig~\ref{fig:ft_phase}).
    For both signals, the color intensity corresponds to scale going from a reference value of \SI{0}{\deci\bel}
    (corresponding to the most intense value of this signal) to \SI{-80}{\deci\bel}.
    The filled lines represent the dispersion relations of transverse (red) and longitudinal (blue) waves.
    Both graphs were obtained using zero-padding of the experimental signal, and Chebychev windowing to mitigate spectral leakage.
    \\ (left) Power spectrum of $\hat{z}(k, \omega)$.
    The blue dotted lines circle the spots where the power spectrum of $\hat{w}(k, \omega)$ is concentrated (see right plot).
    \\ (right) Power spectrum of $\hat{w}(k, \omega)$.
    The red dotted lines circle the spots where the power spectrum of $\hat{z}(k, \omega)$ is concentrated (see right plot).
    }
    \label{fig:ft_nonlinear}
\end{figure}

The evolution of the rivulet geometry due to the instability is at first sight quite complex.
Indeed, the pattern generated is the sum of three contributions of different nature (transverse or longitudinal),
and of different spatial and temporal characteristics.
This can be seen by looking at figures~\ref{fig:spatio_z} and~\ref{fig:spatio_w} which are rather busy and need to time to be properly understood.
One way to disentangle all the different contributions of the pattern,
is to represent the experimental data in the frequency space instead of the real space.
We present on figures~\ref{fig:ft_phase} and~\ref{fig:ft_nonlinear} such a representation,
where the Fourier transform of the position and width of the rivulet are shown as a function of time and space frequencies.
Since each signal is approximately periodic,
it is represented in the Fourier space by a pair of localised dots
that are symmetric with respect to the $(\omega,k)=(0,0)$ point.
This allows for a concise representation of the data :
each wave constituting the pattern correspond to a dot
(or pair of dots for functions only of time)
on these figures.

On figure~\ref{fig:ft_phase}~(left) one can see that the signals correspond to the three different waves forming the instability.
The two red dots aligned on the $k=0$ vertical line
represent the two frequency components of the linear response of the path to the forcing
$\Fampdim e^{i \omega_0 t} + \conjugate{\Fampdim} e^{-i \omega_0 t}$.
The red dot at $k\neq 0$ corresponds to the transverse wave deforming the path,
which is placed along the $\omega_z^-(k)$ dispersion relation, it corresponds to the $\Zampdim e^{i (\omega_z^- t - k_z x)} + \cc$ signal.
The blue dot corresponds to the longitudinal wave modulating the width signal,
it is placed along the $\omega_w^+(k)$ dispersion relation, it corresponds to the $\Wampdim e^{i (\omega_w^+ t - k_w x)} + \cc$ signal.
On figure~\ref{fig:ft_phase}~(right), one can see the same dots as on the left plot,
with the same color intensity but with different colors.
On this plot, the colours now represent the phase of the signals,
they correspond to the argument of $\Fampdim$ (dots on $k = 0$), of $\Zampdim$ (dot on the transverse waves dispersion relation),
and of $\Wampdim$ (dot on the longitudinal waves dispersion relation).

Just by looking at figure~\ref{fig:ft_phase},
one can directly verify the fundamental properties of the \revisecross{pearl necklace }instability.
The pattern is formed of three contributions :
a spatially homogeneous ($k=0$) response to the forcing, a transverse wave and a longitudinal wave.
Both propagative waves are aligned on the same vertical line,
which correspond the~\eqref{eq:rescond_k} condition for resonant interaction,
and their vertical distance correspond to the forcing frequency $\omega_0/(2\pi) = \SI{40}{\hertz}$:
this is nothing more than condition~\eqref{eq:rescond_w}.
With this representation in the dual space, the resonance condition hence take a simple, visual meaning.

Using the Fourier representation, we are able to plot find both the amplitudes and phase of the relevant signals.
By computing the energetic content of the different peaks showed on figure~\ref{fig:ft_phase} (left),
we are able to measure the amplitudes $\abs{\Fampdim}$, $\abs{\Zampdim}$ and $\abs{\Wampdim}$.
These results of these measurements are shown, for varying forcing amplitude,
in figures~\ref{fig:modestructure} (bottom) and~\ref{fig:saturation} (bottom).
But the Fourier representation also gives us access to the absolute phase of each signal.
This is what is shown in figure~\ref{fig:ft_phase} (right): The colored dots correspond to the same signals as on the left plot,
but the colors now represent the phase of the signal.
We are thus able to visually check that the condition~\ref{eq:argrelWZF} on the pattern structure is verified:
on figure~\ref{fig:ft_phase} we see that $\arg(\Fampdim) \approx -\pi/2$, $\arg(\Zampdim)\approx 0$ and thus $\arg(\Wampdim) = -\pi/2 + 0 -\pi/2 = -\pi$.
Both the amplitude and phase of the Fourier spectra can thus be quantitatively linked
with mathematical properties of the pattern derived from our model.

On figure~\ref{fig:ft_nonlinear} we plotted the power spectra of the same signals as in figure~\ref{fig:ft_phase},
while lowering the fade to white threshold.
This allows us to see weaker signals which correspond to secondary modes which are linearly damped.
These modes are excited by nonlinear interactions between transverse displacement $z$ (represented by red arrows),
and longitudinal modulations $w$ (blue arrows).
Due to the $z\rightarrow -z$ mirror symmetry of the system,
the evolution equation for $w$ can display auto-coupled terms proportional to $z^2$ and $w^2$ but never cross coupling terms $zw$.
For the same reason, the evolution equation of $z$ can only display cross coupled terms of the form $wz$.
This symmetry condition thus selects the nonlinear interactions between modes,
and completely explains the spectra observed on figure~\ref{fig:ft_nonlinear},
as indicated by the coloured arrows which represent the contributions responsible for the signals we see.
Note that the absence of signal in $z$ at coordinates $(\omega=\pm 2 \omega_0, k=0)$ on fig.~\ref{fig:ft_nonlinear} (left)
shows that the excitation is sinusoidal in time, with no higher harmonics to be seen.

Using the Fourier representation thus allows one to both make accurate measurements,
and grasp visually several key characteristics of the system and their consequences,
such as the resonance condition,
the pattern structure,
or the symmetry underlining the possible interactions between different waves.

\subsection{Summary, limits and perspectives}
\label{subsec:perspectives}

We have studied the dynamics of liquid rivulets in a Hele-Shaw cell subject to additive homogeneous acoustic forcing.
We established that two types of perturbations can propagate on the rivulet:
deformations of the path and modulations of the width correspond to transverse and longitudinal waves, respectfully.
Each type of wave can evolve at two different phase velocities, corresponding to distinct propagation modes.
These waves are linearly independent, and exponentially attenuated in the absence of forcing.
When imposing a homogeneous additive forcing to the rivulet,
its linear response consists in a homogeneous transverse movement at the same frequency.
This movement acts as a coupling between transverse and longitudinal waves, which can grow by amplifying each other
if an algebraic relationship between the characteristics of the waves and the forcing is verified.
Above a certain threshold, this cooperative interaction between the waves overwhelms the damping and a pattern-forming instability develops.
We are able to understand this threshold and quantitatively retrieve its frequency-dependence.
Our model allows us to develop a fine understanding of the structure of this pattern,
in particular of the amplitude ratio and relative phase between longitudinal and transverse waves.
Nonlinear developments explain the spatial detuning observed in the measurements,
and adding a heuristic term compatible with the symmetries of the system allows us to explain the saturation of the pattern amplitude.
We present data and a quantitative argument to explain the rivulet breakup at high forcing amplitude.
Last, we showed how we can recover almost all the characteristics of the pattern by representing our data in the Fourier space,
which helps visualising in a simple and concise manner how the the experimental system respects the mathematical constrains that our model predicts.

There are several ways one could improve upon the results we present in this study.
The main drawback of the experimental system we use is the finite size of the acoustic excitation.
Being able to move the rivulet homogeneously over a larger spatial extent
would improve the measurement precision on the amplitudes
$\abs{\Zampdim}$, $\abs{\Fampdim}$ and $\abs{\Wampdim}$ as well as on the
wavenumbers $k_z$ and $k_w$. It would also presumably extend the observation window
for this instability to lower frequencies. 
This is feasible by using several loudspeakers, in phase with each other, on each side of the cell.
The membranes of the loudspeakers we use also have a limited range of movement over which the speaker response is linear,
limiting the range of excitation amplitude we can generate.
Using pressurized air nozzles would allow for a greater range of pressures to be explored,
allowing for example the instability to develop at lower excitation frequencies.
On the theoretical side,
the discrepancy of a factor $2.05\pm0.10$ between the measured and predicted prefactor for the instability threshold power law presented on section~\ref{subsec:threshold}
might be solved by a careful rederivation of our result, maybe using a different method.

Let us finish by presenting three possible directions for future research that we think are both attainable,
in the sense that one would only need to make small modifications to the setup in order to observe them,
and, according to our experience with this system, possibly fruitful.

First, one might remark by looking at figures~\ref{fig:ft_phase} and~\ref{fig:ft_nonlinear} that the signal for $\Zampdim$
seems to be on the $\omega = 0$ horizontal line.
This could indicate a frequency locking phenomenon : for a certain range of parameters,
the system strays away from the linear dispersion relations in order to verify $(\omega_z, \omega_w) = (0, \omega_0))$.
This is evidenced by the fact that as shown on the inset of fig.~3 of~\citet{lelay2025},
the time frequency of longitudinal waves $\omega_z$ seems to be zero under a certain threshold.
The physical origin of this frequency locking phenomenon,
which would break the galilean invariance of the amplification mechanism presented on section~\ref{subsec:physics},
is unknown.

Second, there are points in the parameters space that are not considered in this study but seem particularly worthwhile exploring,
possibly leading to new behaviours to explain.
At high forcing amplitude, one might expect secondary instabilities to take place, maybe inducing a drift of the pattern.
For example, the $(\omega=2\,\omega_w, k=2\,k_w)$ mode is naturally forced by the system, and it lies (approximately) on the $\omega_w(k_w)$ dispersion relation : it can be resonantly forced and lead to a secondary instability.
More generally, our system sustains a deep analogy with the Faraday instability,
which is now well understood, and thus one can hope to apply the rich phenomenology of Faraday waves to fluid rivulets.
Using analytical, numerical and experimental techniques that have proven useful for the study of Faraday waves
could be a powerful way to explore and understand the dynamics of the rivulet.
Note that in order to explore much higher forcing amplitude, one will need to make the rivulet more resistant to breakup.
This can be done either (a) by using a thicker rivulet without
augmenting the flow rate, which can be done by diminishing the cell
spacing $b$ and/or increasing the fluid viscosity,
or (b) by avoiding the rupture when $w$ reaches zero ---i.e.\ in the
limit where the rivulet is a film linking Plateau borders on the
walls~\citep{drenckhan2007}--- to explore ``negative'' width $w$ solutions, which can be done by replacing the oil by a water-surfactant mix.

Last, using an improved setup,
it could be very interesting to use an acoustic forcing with different spatiotemporal characteristics than the one used in this study
(monochromatic excitation at $(\omega=\pm\omega_0, k= 0)$).
By modulating the forcing both in time and space, for example by using a forcing with two or more frequency components,
one could excite several waves simultaneously and observe their interactions.
By using a forcing with a continuous frequency spectrum,
one would be able to observe a continuum of waves that interact nonlinearly,
with possible energy exchange and cascading between scales.
The result could be accurately described by the quasi-1D wave turbulence theory,
which have already been observed experimentally for gravity-capillary waves~\citep{ricard2021}.

For the purpose of open access, the author has applied a Creative Commons Attribution (CC BY) license
to any Author Accepted Manuscript version arising from this submission.

\backsection[Supplementary movies]{\label{SupMat}Supplementary movies are available at [...]}

\backsection[Acknowledgements]{G. L. is deeply grateful to François Gallaire, François Pétrélis and Michael Berhanu for insightful discussions and for the helpful advices they provided.}

\backsection[Founding]{This research received special funds from laboratory MSC (UMR 7057).}

\backsection[Declaration of interests]{The authors report no conflict of interest.}

\backsection[Data availability statement]{The data that support the findings of this study are available from the corresponding author upon reasonable request.}

\backsection[Author ORCIDs]{G. Le Lay, https://orcid.org/0009-0002-2654-368X ; A. Daerr, https://orcid.org/0000-0002-1083-9681}

\clearpage
\appendix
\section{Nonlinear development of the dynamical equations}\label{sec:appa}

In this section, we develop the dynamical equations of the system up to third order in $\eps$.

We define $\theta$ as the angle the rivulet makes with the vertical, and geometrically we have, up to third order in $\eps$,
\begin{align*}
    \sin\theta &= \frac{\partial_x z}{\sqrt{1 + (\partial_x z)^2}}
    \\
    &= \eps\partial_x z_1 + \eps^2\partial_x z_2 + \eps^3\qty(\partial_x z_3 - \frac12(\partial_x z_1)^3) + o(\eps^4)
    \\
    \qqtext{and} \cos\theta &= \frac{1}{\sqrt{1 + (\partial_x z)^2}}
    \\
    &= 1 \Cgeo{- \eps^2 \frac12 \qty(\partial_x z_1)^2} - \eps^3 \partial_x z_1\partial_x z_2 + o(\eps^4) \ .
\end{align*}

Our strategy to obtain the nonlinear equations is the following:
it seems intuitive at first to write the nonlinear equations for the behaviour of the rivulet in curvilinear coordinates,
projecting all the vectors unto a local Frenet–Serret basis.
This seems to allow the natural decoupling between streamwise forces (such as the bulk viscosity) and normal forces
(such as pressure variations between the sides of the rivulet).
However since the rivulet, and thus the vector basis, moves with both space and time,
any definition of a curvilinear parameter is ambiguous,
and the writing of the dynamical equations in a frame of reference moving both in space and time quickly becomes extremely tedious.
A simpler alternative is to introduce the streamwise and normal components of the speed $u_s = u\cos\theta + v\sin\theta$ and $u_n = -u \sin\theta + v\cos\theta$.
We then write the equations using these variables, but projecting them onto the familiar, static basis $(\hat{\vb{x}}, \hat{\vb{z}})$:
\begin{align}
    w(\partial_t + u_s\partial_s) v
    =& - w\mu v
    +  \Cgat{w \Gamma \partial_s(\kappa_w)} \sin\theta
    + \qty(\Gamma \kappa_z - \mucl u_n ) \cos\theta
    \\
    w(\partial_t + u_s\partial_s) u
    =& w g - w\mu u
    + \Cgat{ w \Gamma \partial_s(\kappa_w)} \cos\theta
    + \qty(\Gamma \kappa_z - \mucl u_n )(-\sin\theta) 
    \\
    (\partial_t + u_s\partial_s) w =& -w \partial_s u_s
\end{align}
with $\partial_s =\cos\theta \partial_x$.
The equations thus take a simple form,
but there is still a lot of work to be done in order to obtain the approximate evolution equations for our chosen variables, $z$ and $w$.

In order to abbreviate the forthcoming equations,
we will use the following notation for the time derivative in the advected frame of reference:
$\adv = \partial_t + u_0 \partial_x$

\subsection{Preliminary computations}

\subsubsection{Normal velocity $\un$}
The kinematic advection condition imposes
\begin{align*}
    \vb{u}\cdot\vb{e}_z = v &= \partial_t z  + u_s\partial_s z
\end{align*}
We use $v = u_s\sin\theta + u_n\cos\theta$, $\sin\theta = \frac{\partial_x z}{\sqrt{1+(\partial_x z)^2}} = z_x \cos \theta$, and $\partial_s \bullet = \cos\theta \partial_x \bullet$
\begin{align*}
    u_s\sin\theta + u_n\cos\theta &= \partial_t z  + u_s\cos\theta \partial_x z \\
    u_s\partial_x z\cos\theta + u_n\cos\theta &= \partial_t z  + u_s\cos\theta \partial_x z
\end{align*}
Hence
\begin{align}
    u_n\cos\theta &= \partial_t z
\end{align}
thus $\un_1 = \partial_t z_1$, $\un_2 = \partial_t z_2$, and $\un_3 = \partial_t z_3 + \partial_\Tdim z_1 \Cgeo{+ \frac12(\partial_x z_1)^2\partial_t z_1 }$ (anticipating that $\partial_x z_2=0$).

\subsubsection{Streamwise velocity $\us$}
By definition
\begin{align}
    u_s &= u\cos\theta+ v \sin\theta  \\
    \us_0 &= u_0 \\
    \us_1 &= u_1
    \\ 
    \us_2 &= u_2 \Cgeo{-u_0\frac12 (\partial_x z_1)^2} + v_1\partial_x z_1
    \nonumber
    \\
    &= u_2 + \Cgeo{\frac12}  u_0 (\partial_x z_1)^2 + \partial_t{z_1}\partial_{x}z_1
    \\ 
    \us_3 &= u_3\Cgeo{ -u_1\frac12 (\partial_x z_1)^2 -u_0 \partial_x z_1\partial_x z_2}
    +v_1\partial_x z_2 + v_2\partial_x z_1 \nonumber
    \\
    &= u_3  +u_1\frac12 (\partial_x z_1)^2
    +u_0\partial_x z_1\partial_x z_2  +\partial_t z_1\partial_x z_2+\partial_t z_2\partial_x z_1
\end{align}

\subsubsection{Link between $v$ and $z$}
By definition
\begin{align}
    v &= \un\cos\theta + u_s \sin\theta
    \\
    \text{i.e.}\quad v_1 &= \un_1 + \us_0\partial_x z_1 = (\partial_t + u_0\partial_x)z_1
    \label{eq:annexv1}
    \\
    v_2 &= \un_2 + \us_0\partial_x z_2 \Cgeo{+ \us_1\partial_x z_1} = (\partial_t + u_0\partial_x)z_2 \Cgeo{+ u_1\partial_x z_1}
    \label{eq:annexv2}
    \\
    v_3 &= \un_3 \Cgeo{-\frac12(\partial_x z_1)^2\un_1} + \us_0\partial_x z_3 \Cgeo{+ \us_1\partial_x z_2 + \us_2\partial_x z_1 - \frac16\us_0(\partial_x z_1)^3}
    \nonumber
    \\
    &= \partial_\Tdim z_1 + (\partial_t + u_0\partial_x)z_3 \Cgeo{+ u_1\partial_x z_2 + u_2\partial_x z_1 + \partial_t z_1 (\partial_x z_1)^2 + \frac13 u_0(\partial_x z_1)^3}
    \label{eq:annexv3}
\end{align}

\subsubsection{Curvatures}
The effective curvatures of the path $\kappa_z$ and width profile $\kappa_w$ correspond,
up to second order in $\eps$, to the second space derivative of the respective variables:
\begin{align*}
    \kappa_{z, 1} =& \partial_{xx}z_1 & \kappa_{w, 1} =& \partial_{xx}w_1 \\
    \kappa_{z, 2} =& \partial_{xx}z_2 & \kappa_{w, 2} =& \partial_{xx}w_2
\end{align*}
Going to the third order, we face a difficulty : the derivation of \citet{parkHomsy84} is only valid for linear approximations of the curvatures $\kappa_{z,w}$.
To push to system to third order, we assume that the curvatures take the following form:
\begin{align*}
    \kappa_{z} =&\, \frac12\frac{\partial_{xx}(z + w/2)}{(1+(\partial_x (z + w/2))^2)^{3/2}} + \frac12\frac{\partial_{xx}(z - w/2)}{(1+(\partial_x (z - w/2))^2)^{3/2}}
    \\
    \kappa_{w} =&\, \frac{\partial_{xx}(z + w/2)}{(1+(\partial_x (z + w/2))^2)^{3/2}} - \frac{\partial_{xx}(z - w/2)}{(1+(\partial_x (z - w/2))^2)^{3/2}} .
\end{align*}

This assumption leads to the following third order development:
\begin{align*}
    \kappa_{z, 3} =& \partial_{xx}z_3 \Cgeo{-\frac32 \qty[(\partial_x z_1)^2+(\partial_x w_1)^2/4]\partial_{xx}z_1 - \frac34 (\partial_x z_1)(\partial_x w_1)\partial_{xx}w_1}
    \\
    \kappa_{w, 3} =& \partial_{xx}w_3 \Cgeo{-\frac32 \qty[(\partial_x z_1)^2+(\partial_x w_1)^2/4]\partial_{xx}w_1 - \frac34 (\partial_x z_1)(\partial_x w_1)\partial_{xx}z_1}
\end{align*}

\subsubsection{Mass conservation}
Mass conservation states
\begin{align}
    \partial_t w =& -\partial_s (w\,\us)
    \\
    (\partial_t + u_0\partial_x) w_1 =& -w_0\partial_x u_1
    \label{eq:massconsw1}
    \\
    (\partial_t + u_0\partial_x) w_2 =& -w_0\partial_x \us_2  - \partial_x(w_1 u_1) \label{eq:massconsw2}
    \\
    =& -w_0\partial_x u_2  - \partial_x(w_1 u_1) \Cgeo{-\frac12 u_0 w_0\partial_x (\partial_x z_1)^2} - \Cgeo{w_0\partial_x \qty(\partial_t{z_1}\partial_{x}z_1)}
    \nonumber
    \\
    (\partial_t + u_0\partial_x) w_3 + \partial_\Tdim w_1
    =&
    -w_0\partial_x \us_3  - \partial_x(w_2 u_1 +w_1\us_2)
    \Cgeo{+\frac12(\partial_x z_1)^2 \partial_x(w_0 u_1  +u_0 w_1)}
    \nonumber\\
    =&
    -w_0\partial_x u_3 -w_0\partial_x \qty( u_1\frac12 (\partial_x z_1)^2 + u_0\partial_x z_1\partial_x z_2  +\partial_t z_1\partial_x z_2+\partial_t z_2\partial_x z_1 )
    \nonumber\\
    &- \partial_x\qty(w_2 u_1 + w_1 u_2 + w_1\Cgeo{\qty[\frac12 u_0 (\partial_x z_1)^2 + \partial_t{z_1}\partial_{x}z_1]})
    \nonumber\\
    & \Cgeo{+\frac12(\partial_x z_1)^2 \qty(w_0\partial_x u_1  +u_0\partial_x w_1)}
    \label{eq:massconsw3}
\end{align}

\subsubsection{Linear results}

The linear equations can be written (see~\ref{subsec:waves})
\begin{align}
\Lz z_1 &= 0 \label{eq:app:linz}
\\
\Lw w_1 &= 0 \label{eq:app:linw}
\\
(\partial_t + u_0 \partial_x) w_1 &= -w_0\partial_x u_1 \label{eq:app:linu}
\end{align}

From~\eqref{eq:app:linz} we can write $z_1 = \Zampdim e^{i(\omega_z t- k_z x)} + \cc$,
from~\eqref{eq:app:linw} we can write $w_1 = \Wampdim e^{i(\omega_w t- k_w x)} + \cc$,and,
since~\eqref{eq:app:linu} is linear, $u_1 = \Uampdim e^{i(\omega_w t- k_w x)} + \cc$.

The dispersions relations~\eqref{eq:dispersionrelation} can be written in the form
\begin{subequations}
\begin{align}
    \omega_{z} &= (u_0 + \varepsilon_z \vcap) k_z
    \\
    \omega_{w} &= u_0 k_w +\varepsilon_w \vcap w_0 {k_w}^2
\end{align}
\end{subequations}
where $\varepsilon_z$ and $\varepsilon_w$ can be +1 or -1 depending on the branch that we consider.

Last, from~\eqref{eq:app:linu} we obtain
\begin{align}
    (i\omega_w -i u_0 k_w) \Wampdim &= i w_0 k_w \Uampdim \nonumber \\
    \Uampdim &= \frac{\omega_w - u_0 k_w}{w_0 k_w}\Wampdim \nonumber\\
    \Uampdim &= \varepsilon_w \vcap k_w\Wampdim.
\end{align}

\clearpage
\subsection{Second order in $\eps$ evolution equations} \label{subsec:annex:order2}

\subsubsection{Second order evolution equation for $z$}
The dynamical equation projected onto $z$, at second order, reads
\begin{align}
    w_0\adv  v_2 \quad& \nonumber
    \\
    + (w_1\adv + w_0 u_1\partial_x ) v_1
    =
    & \Cgat{ w_0\Gamma \partial_{x} \kappa_{w, 1}(\partial_x z_1)} \nonumber
    \\
    & + \Gamma \kappa_{z, 2} + \Pi(t)
\end{align}
which leads to
\begin{align}
    w_0 \Lz z_2
    +w_0 (\partial_x z_1) (\adv u_1)+ 2 w_0 u_1  \adv \partial_x z_1 + w_1{\adv}^2 z_1
    =
    & \Cgat{ w_0 \Gamma \partial_x z_1 \partial_{xxx} w_1} + \Pi(t)
\end{align}
which correspond to equation~\eqref{eq:secondorder_z} with
\begin{align}
    \NLzz(z_1, w_1, u_1)
    =& { - (\partial_x z_1 )(\adv u_1) - 2 u_1  \adv\partial_x z_1) }
    \nonumber \\
    & { - \frac{w_1}{w_0}{\adv}^2 z_1 + \Gamma \partial_x z_1 \partial_{xxx} w_1 }
\end{align}

\subsubsection{Order 2 in \revise{$u$ and } $w$}

The dynamical equation projected onto $w$, at second order, is
\begin{align}
    w_0\adv  u_2\quad& \nonumber
    \\
    + (w_1\adv + w_0 u_1\partial_x ) u_1
    =& \Cgat{w_0\Gamma \partial_{x} \kappa_{w, 2}}  \nonumber
    \\
    & \Cgat{+ w_1\Gamma \partial_{x} \kappa_{w, 1}} \nonumber
    \\
    & - \Gamma \kappa_{z, 1}(\partial_x z_1)
    \label{eq:annex:order2wstart}
\end{align}
on which we need to take the space derivative $-\partial_x$\eqref{eq:annex:order2wstart}
\begin{align}
    \adv  (-w_0\partial_x u_2)\quad& \nonumber
    \\
    + w_1\adv (-\partial_x u_1) + u_1\partial_x (- w_0\partial_x  u_1 ) + (\partial_x  u_1)(-w_0 \partial_x u_1)
    =
    & \Cgat{w_0\Gamma \partial_{xxxx} w_2 } \nonumber
    \\
    & \Cgat{- \Gamma \partial_x \qty[ w_1 \partial_{xxx} w_1] } \nonumber
    \\
    & + \Gamma \partial_x \qty[ (\partial_{xx}z_1)(\partial_x z_1) ]
\end{align}
after that we can replace using the appropriate expressions and obtain
\begin{align}
    \Lw w_2 \quad& \nonumber
    \\
    + \adv \partial_x (w_1 u_1) + w_0 \adv \partial_x\qty[ (\partial_x z_1)\qty(\qty(\partial_t - \frac12 u_0) z_1) ] \quad& \nonumber
    \\
    + \frac{w_1}{w_0}{\adv}^2 w_1 + (\partial_x  u_1 + u_1\partial_x)(\adv w_1)
    =
    & \Cgat{- \Gamma \partial_x \qty[ w_1 \partial_{xxx} w_1] } \nonumber
    \\
    & + \Gamma \partial_x \qty[ (\partial_{xx}z_1)(\partial_x z_1) ]
\end{align}
from which the definition of $\NLww(z_1, w_1, u_1)$ in equation~\eqref{eq:secondorder_w} is immediate.

\clearpage
\subsection{Third order in $\eps$ evolution equations}  \label{subsec:annex:order3}

\subsubsection{Third order evolution equation for $z$}
The dynamical equation projected $e_z$ reads
\begin{align}
    w_0\adv  v_3 + w_0\partial_\Tdim v_1 \quad& \nonumber
    \\
    + (w_1\adv + w_0 u_1\partial_x ) v_2 \quad& \nonumber
    \\
    + (w_2\adv + w_0 \us_2\partial_x + w_1 u_1\partial_x) v_1  \quad& \nonumber
    \\
    \Cgeo{-\frac12 (\partial_x z_1)^2}w_0 u_0\partial_x v_1
    =&
    -w_0 \mu v_1 - \mucl \partial_t z_1\nonumber
    \\
    & \Cgat{+ w_0\Gamma \partial_{x} \kappa_{w, 2}(\partial_x z_1)}
    \Cgat{+ w_0\Gamma \partial_{x} \kappa_{w, 1}(\partial_x z_2)} \nonumber
    \\
    & \Cgat{+ w_1 \Gamma \partial_{x} \kappa_{w, 1}(\partial_x z_1)} \nonumber
    \\
    & + \Gamma \kappa_{z, 3} - \Cgeo{\frac12 (\partial_x z_1)^2\Gamma \kappa_{z, 1}}
\end{align}
which, after replacing using equations~\eqref{eq:annexv1},~\eqref{eq:annexv2} and~\eqref{eq:annexv3}, leads to
\begin{align}
    w_0\adv  (\partial_\Tdim z_1 + \adv z_3) + w_0\partial_\Tdim \adv z_1 \quad& \nonumber
    \\
    +w_0 \adv \qty(u_1\partial_x z_2 + u_2\partial_x z_1 + \partial_t z_1(\partial_x z_1)^2 + \frac13 u_0 (\partial_x z_1)^3) \quad& \nonumber
    \\
    + (w_1\adv + w_0 u_1\partial_x ) (\adv z_2 + u_1\partial_x z_1) \quad& \nonumber
    \\
    + (w_2\adv + w_0 \qty(u_2 + \frac12 u_0(\partial_x z_1)^2 + \partial_t z_1 \partial_x z_1)\partial_x  \quad& \nonumber
    \\
    + w_1 u_1\partial_x) \adv z_1 \quad& \nonumber
    \\
    \Cgeo{-\frac12 (\partial_x z_1)^2}w_0 u_0\partial_x \adv z_1
    =&
    -w_0 \mu \adv z_1 - \mucl \partial_t z_1 \nonumber
    \\
    & {+ w_0\Gamma \partial_{xxx}w_2(\partial_x z_1)} \nonumber
    \\
    & {+ w_0\Gamma \partial_{xxx}w_1(\partial_x z_2)} \nonumber
    \\
    & {+ w_1 \Gamma \partial_{xxx}w_1 (\partial_x z_1)} \nonumber
    \\
    & + \Gamma \partial_{xx}z_3 - \Cgeo{\frac12 (\partial_x z_1)^2\Gamma \partial_{xx}z_1} \nonumber
    \\
    & -\frac32 \Gamma \qty[(\partial_x z_1)^2+(\partial_x w_1)^2/4]\partial_{xx}z_1 \nonumber
    \\
    & - \frac34 \Gamma(\partial_x z_1)(\partial_x w_1)\partial_{xx}w_1 .
\end{align}

Using the properties of $z_2$, $w_2$ and $u_2$ one obtains
\begin{align}
    w_0 \Lz z_3 + 2 w_0\partial_\Tdim \adv z_1  \quad& \nonumber
    \\
    +w_0 \adv \qty( \partial_t z_1(\partial_x z_1)^2 ) + \frac13 u_0 w_0 \adv (\partial_x z_1)^3 \quad& \nonumber
    \\
    + w_1 \partial_{tt} z_2 + (w_1\adv + w_0 u_1\partial_x) (u_1\partial_x z_1) \quad& \nonumber
    \\
    + (w_0\partial_t z_1 \partial_x z_1 + w_1 u_1) \partial_x \adv z_1
    =&
    -w_0 \mu \adv z_1 - \mucl \partial_t z_1
    \\
    & \Cgat{+ w_1 \Gamma \partial_{xxx}w_1 (\partial_x z_1)} \nonumber
    \\
    & - \Cgeo{\frac12 (\partial_x z_1)^2\Gamma \partial_{xx}z_1} \nonumber
    \\
    & -\frac32 \Gamma \qty[(\partial_x z_1)^2+(\partial_x w_1)^2/4]\partial_{xx}z_1 \nonumber
    \\
    & - \frac34 \Gamma (\partial_x z_1)(\partial_x w_1)\partial_{xx}w_1 .
\end{align}
And considering only the terms susceptible to resonate, we write:
\begin{align}
    w_0 \Lz z_3 + 2 w_0\partial_\Tdim \adv z_1  \quad& \nonumber
    \\
    +w_0 \adv \qty( \partial_t z_1(\partial_x z_1)^2 ) + \frac13 w_0 u_0 \adv (\partial_x z_1)^3 \quad& \nonumber
    \\
    + 2w_1 u_1\adv\partial_x z_1 + w_0 {u_1}^2\partial_{xx}z_1  \quad& \nonumber
    \\
    + w_0\partial_t z_1 \partial_x z_1  \partial_x \adv z_1
    =&
    -w_0 \mu \adv z_1 - \mucl \partial_t z_1 - w_1 \partial_{tt} z_2 \nonumber
    \\
    & - \Cgeo{\frac12 (\partial_x z_1)^2\Gamma \partial_{xx}z_1} \nonumber
    \\
    & -\frac32 \qty[(\partial_x z_1)^2+(\partial_x w_1)^2/4]\partial_{xx}z_1 \nonumber
    \\
    & - \frac34 (\partial_x z_1)(\partial_x w_1)\partial_{xx}w_1 .
\end{align}
We now adopt use the complex expressions of our variables,
we write the phase speed of transverse waves as $v_z = \omega_z / k_z$,
and the solvability condition becomes:

\begin{align}
    2 w_0 \qty[i k_z(v_z - u_0)]\partial_\Tdim \Zampdim  \quad& \nonumber
    \\
    -3 i w_0 v_z {k_z}^3 \qty[i k_z(v_z - u_0)] \abs{\Zampdim}^2 \Zampdim \quad& \nonumber
    \\
    -  i w_0 u_0 {k_z}^3 \qty[i k_z(v_z - u_0)]  \abs{\Zampdim}^2 \Zampdim \quad& \nonumber
    \\
    - 2i k_z \qty[i k_z(v_z - u_0)] (\Wampdim \conjugate{\Uampdim} + \conjugate{\Wampdim}\Uampdim) \Zampdim \quad& \nonumber
    \\
    - 2 w_0 {k_z}^2 \abs{\Uampdim}^2\Zampdim \quad& \nonumber
    \\
    - 3i w_0 v_z {k_z}^3\qty[i k_z(v_z - u_0)] \abs{\Zampdim}^2 \Zampdim
    =&
    -w_0 \mu \qty[i k_z(v_z - u_0)] \Zampdim - i \mucl v_z k_z \Zampdim + {\omega_0}^2 \Wampdim\conjugate{\Fampdim} \nonumber
    \\
    & - \frac32 w_0 \vcap^2 {k_z}^4 \abs{\Zampdim}^2 \Zampdim \nonumber
    \\
    & -\frac92 w_0 \vcap^2 {k_z}^4 \abs{\Zampdim}^2 \Zampdim -\frac34 w_0 \vcap^2 {k_z}^2 {k_w}^2 \abs{\Wampdim}^2\Zampdim \nonumber
    \\
    & - \frac32 w_0 \vcap^2 {k_z}{k_w}^3 \abs{\Wampdim}^2 \Zampdim.
\end{align}

and after some reordering:

\begin{align}
    2 \partial_\Tdim \Zampdim
    =&
    -\mu \qty( 1- \frac{\mucl}{w_0 \mu} \frac{v_z}{u_0 - v_z}) \Zampdim + i\frac{1}{1-v_z/u_0} \frac{{\omega_0}^2}{w_0 u_0 k_z} \Wampdim\conjugate{\Fampdim} \nonumber
    \\
    & + i {k_z}^3 \qty(u_0 + 6 v_z + 6\frac{\vcap^2}{v_z - u_0}) \abs{\Zampdim}^2 \Zampdim \nonumber
    \\
    & + i  {k_w}^2 \qty(\qty(\frac32 k_w - \frac14 k_z) \frac{\vcap^2}{v_z - u_0} + 4 \varepsilon_w  \frac{k_{z}}{k_w} \frac{\vcap}{w_0}) \abs{\Wampdim}^2 \Zampdim .
\end{align}
Now using that $v_z = u_0 + \varepsilon_z \vcap$, and anticipating that $k_z=k_w = k$, we can obtain
\begin{align}
    2 \partial_\Tdim \Zampdim
    =&
    -\mu \qty( 1 + \varepsilon_z \frac{\mucl}{w_0 \mu} \qty(\frac{u_0}{\vcap} + \varepsilon_z)) \Zampdim - i \varepsilon_z \frac{{\omega_0}^2}{w_0 \vcap k} \Wampdim\conjugate{\Fampdim} \nonumber
    \\
    & + 7 i {k}^3 u_0 \abs{\Zampdim}^2 \Zampdim \nonumber
    \\
    & + i  {k}^2 \vcap \qty(- \varepsilon_z \frac54 k + \frac{4 \varepsilon_w}{w_0}) \abs{\Wampdim}^2 \Zampdim .
\end{align}

\clearpage
\subsubsection{Third order evolution equation for $w$}
The dynamical equation projected $e_x$ reads
\begin{align}
    w_0\adv  u_3 + w_0 \partial_\Tdim u_1\quad& \nonumber
    \\
    + (w_1\adv + w_0 u_1\partial_x ) u_2 \quad& \nonumber
    \\
    + (w_2\adv + w_0 \us_2\partial_x + w_1 u_1\partial_x) u_1  \quad& \nonumber
    \\
    \Cgeo{-\frac12 (\partial_x z_1)^2}w_0 u_0\partial_x u_1
    =&
    -w_0 \mu u_1 \nonumber
    \\
    & \Cgat{+ w_0\Gamma \partial_{x} \kappa_{w, 3}}  \nonumber
    \\
    & \Cgat{+ w_1\Gamma \partial_{x} \kappa_{w, 2}}
    \Cgat{+ w_2\Gamma \partial_{x} \kappa_{w, 1}} \nonumber
    \\
    & \Cgat{- w_0\Gamma \partial_{x} \kappa_{w, 1}} (\partial_x z_1)^2 \nonumber
    \\
    & - \Gamma \kappa_{z, 1}(\partial_x z_2) - \Gamma \kappa_{z, 2}(\partial_x z_1) \nonumber
    \\
    & - \Pi (\partial_x z_1) .
    \label{eq:annex:order3wstart}
\end{align}
We then use that $w_0 \adv^2 z_2 = \Pi$ and compute $-\partial_x$\eqref{eq:annex:order3wstart}
\begin{align}
    \adv  (-w_0\partial_x u_3)  + \partial_\Tdim (-w_0\partial_x u_1)\quad& \nonumber
    \\
    -\partial_x w_1 \partial_t u_2 + \frac{w_1}{w_0}\partial_t (-w_0\partial_x u_2) \quad& \nonumber
    \\
    + \partial_x \qty(\frac{u_0}{w_0}w_1 + u_1 ) (-w_0\partial_x u_2) \quad& \nonumber
    \\
    -\partial_x w_2 \partial_t u_1 + \frac{w_2}{w_0}\partial_t (-w_0\partial_x u_1) \quad& \nonumber
    \\
    + \partial_x \qty(\frac{u_0}{w_0}w_2 + \us_2 + \frac{1}{w_0}w_1 u_1) (-w_0\partial_x u_1)  \quad& \nonumber
    \\
    \Cgeo{-\frac12 u_0 } \partial_x \qty( (\partial_x z_1)^2 (-w_0\partial_x u_1) )
    =&
    - \mu (-w_0\partial_x u_1)
    \\
    & \Cgat{- w_0\Gamma \partial_{xxxx} w_3} + \frac32 w_0\Gamma \partial_{xx}\qty((\partial_x w_1)^2\partial_{xx}w_1) \nonumber
    \\
    & \Cgat{- \Gamma \partial_x\qty( w_1 \partial_{xxx} w_2
    + w_2 \partial_{xxx} w_1)} \nonumber
    \\
    & + w_0\Gamma  \partial_x \qty( (\partial_{xxx} w_1) (\partial_x z_1)^2) \nonumber
    \\
    & + \Gamma \partial_x\qty((\partial_{xx} z_1)(\partial_x z_2)) + \Gamma \partial_x\qty((\partial_{xx} z_2)(\partial_x z_1)) \nonumber
    \\
    & + w_0 \partial_x\qty( \adv^2 z_2  (\partial_x z_1)) \nonumber
\end{align}
We now have to use the mass conservation equations~\eqref{eq:massconsw1} and~\eqref{eq:massconsw3} in order to obtain
\small
\begin{align}
    \adv  (\adv w_3 + \partial_\Tdim w_1)  + \partial_\Tdim \adv w_1\quad& \nonumber
    \\
    + w_0 \adv \partial_x\qty( u_1\frac12 (\partial_x z_1)^2 + u_0\partial_x z_1\partial_x z_2  +\partial_t z_1\partial_x z_2+\partial_t z_2\partial_x z_1 ) \quad& \nonumber
    \\
    + \adv \partial_x\qty(w_2 u_1 + w_1 u_2 + w_1\Cgeo{\qty[\frac12 u_0 (\partial_x z_1)^2 + \partial_t{z_1}\partial_{x}z_1]})  \quad& \nonumber
    \\
    - \adv \qty(\Cgeo{\frac12(\partial_x z_1)^2 \qty(w_0\partial_x u_1  +u_0\partial_x w_1)})  \quad& \nonumber
    \\
    -\partial_x w_1 \partial_t u_2 + \frac{w_1}{w_0}\partial_t (-w_0\partial_x u_2) \quad& \nonumber
    \\
    + \partial_x \qty(\frac{u_0}{w_0}w_1 + u_1 ) (-w_0\partial_x u_2) \quad& \nonumber
    \\
    -\partial_x w_2 \partial_t u_1 + \frac{w_2}{w_0}\partial_t \adv w_1 \quad& \nonumber
    \\
    + \partial_x \qty[ \qty(\frac{u_0}{w_0}w_2 + \us_2 + \frac{1}{w_0}w_1 u_1) \adv w_1 ] \quad& \nonumber
    \\
    \Cgeo{-\frac12 u_0 } \partial_x \qty( (\partial_x z_1)^2 \adv w_1 )
    =&
    - \mu \adv w_1 \nonumber
    \\
    & \Cgat{- w_0\Gamma \partial_{xxxx} w_3} + \frac32 w_0\Gamma \partial_{xx}\qty((\partial_x w_1)^2\partial_{xx}w_1) \nonumber
    \\
    & \Cgat{- \Gamma \partial_x\qty( w_1 \partial_{xxx} w_2
    + w_2 \partial_{xxx} w_1)} \nonumber
    \\
    & + w_0\Gamma  \partial_x \qty( (\partial_{xxx} w_1) (\partial_x z_1)^2) \nonumber
    \\
    & + \Gamma \partial_x\qty((\partial_{xx} z_1)(\partial_x z_2)) + \Gamma \partial_x\qty((\partial_{xx} z_2)(\partial_x z_1)) \nonumber
    \\
    & + w_0 \partial_x\qty( \adv^2 z_2  (\partial_x z_1)) .
\end{align}
\normalsize
We now use equation~\eqref{eq:massconsw2} and do some simplifications to obtain
\small
\begin{align}
    \mathcal{L}_w w_3 + 2\adv \partial_\Tdim w_1) \quad& \nonumber
    \\
    + w_0 \adv \partial_x\qty( u_1\frac12 (\partial_x z_1)^2  + \partial_x z_1 \partial_t z_2 ) \quad& \nonumber
    \\
    + \adv \partial_x\qty( w_1\Cgeo{\qty[\frac12 u_0 (\partial_x z_1)^2 + \partial_t{z_1}\partial_{x}z_1]}) \quad& \nonumber
    \\
    - \adv \qty(\Cgeo{\frac12(\partial_x z_1)^2 \qty(-\partial_t w_1)})  \quad& \nonumber
    \\
    + \partial_x \qty[\qty(\frac12 u_0(\partial_x z_1)^2 + \partial_t z_1 \partial_x z_1 + \frac{1}{w_0}w_1 u_1) \adv w_1 ]  \quad& \nonumber
    \\
    \Cgeo{-\frac12 u_0 } \partial_x \qty( (\partial_x z_1)^2 \adv w_1 )
    =&
    - \mu \adv w_1 \nonumber
    \\
    & + \frac32 w_0\Gamma \partial_{xx}\qty((\partial_x w_1)^2\partial_{xx}w_1) \nonumber
    \\
    & + w_0\Gamma  \partial_x \qty( (\partial_{xxx} w_1) (\partial_x z_1)^2) \nonumber
    \\
    & + w_0  \partial_{xx} z_1 \partial_{tt} z_2    .
\end{align}
\normalsize
Finally, we use the complex expressions of $z_1$, $w_1$ and $z_2$, with $\adv w = i (\omega_w - u_0 k_w) = \varepsilon_w i \vcap k^2 w_0$
\begin{align}
(\varepsilon_w i \vcap k^2 w_0) 2 \partial_\Tdim W) \quad& \nonumber
\\
+(\varepsilon_w i \vcap k^2 w_0) (-ik)\qty( u_0 W\frac12 \abs{kZ}^2  + w_0 (-ik) Z (i\omega_0) F ) \quad& \nonumber
\\
+ (\varepsilon_w i \vcap k^2 w_0) (-ik)\qty( W \Cgeo{\qty[\frac12 u_0 \abs{kZ}^2 + v_z \abs{kZ}^2]}) \quad& \nonumber
\\
+ (\varepsilon_w i \vcap k^2 w_0) \qty(\Cgeo{\frac12 \abs{kZ}^2 i\omega_w W})  \quad& \nonumber
\\
+ (-ik) (\varepsilon_w i \vcap k^2 w_0) \qty(\frac12 u_0 \abs{kZ}^2  - v_z \abs{kZ}^2  + \frac{u_0}{{w_0}^2}\abs{W}^2) W   \quad& \nonumber
\\
+\frac12 u_0 (\varepsilon_w i \vcap k^2 w_0) (ik) \abs{kZ}^2  W
=&
- (\varepsilon_w i \vcap k^2 w_0) \mu w_1 \nonumber
\\
& + \frac32 w_0\Gamma k^6 \abs{W}^2 W \nonumber
\\
& + w_0\Gamma  \abs{kZ}^2 k^4 W  \nonumber
\\
& + w_0  k^2 {\omega_0}^2 F Z
\end{align}
And finally obtain
\begin{align}
    2\partial_\Tdim \Wampdim
    =&
    - \mu \Wampdim - i  \varepsilon_w k \omega_0 \qty( \frac{{\omega_0}}{k \vcap} -\varepsilon_w w_0 k )  \Fampdim \Zampdim \nonumber
    \\
    & -\varepsilon_w 2i  w_0 k^2 \vcap \abs{k\Zampdim}^2 \Wampdim + i \qty( k\frac{u_0}{{w_0}^2} -\varepsilon_w \frac32 w_0 \vcap k^4 ) \abs{\Wampdim}^2 \Wampdim
\end{align}
which corresponds to equation~\eqref{eq:order3_W}.

\bibliographystyle{jfm}
\providecommand{\noopsort}[1]{}\providecommand{\singleletter}[1]{#1}%

\end{document}